\documentclass[
  twoside,
  twocolumn,
  prb,
  preprintnumbers,
  floatfix,
  showpacs, superscriptaddress,
  ]{revtex4-1}
\usepackage{amsmath}
\usepackage{amssymb}
\usepackage{color}
\usepackage{graphicx}
\usepackage{tabularx}
\usepackage{dcolumn} 
\usepackage{bm}      
\usepackage{array}   
\usepackage{amsfonts}
\usepackage{siunitx} 
\usepackage{ulem}
\usepackage{soul}
\usepackage{times}
\usepackage{chemformula}

\usepackage{xr-hyper}
\usepackage[implicit=false]{hyperref}

\makeatletter
\newcommand*{\addFileDependency}[1]{
  \typeout{(#1)}%
  \@addtofilelist{#1}%
  \IfFileExists{#1}{}{\typeout{No file #1.}}%
}
\makeatother
\newcommand*{\myexternaldocument}[1]{%
    \externaldocument{#1}%
    \addFileDependency{#1.tex}%
    \addFileDependency{#1.aux}%
}

\myexternaldocument{supplement}

\newcolumntype{.}[1]{D{.}{.}{#1}}

\begin{document}

\title{How Correlated Adsorbate Dynamics on Realistic Substrates Can Give Rise to 1/\textbf{$\omega$} Electric-Field Noise in Surface Ion Traps}

\author{Benjamin Foulon}
\affiliation{Department of Chemical Engineering, Brown University, Providence, RI 02912, USA}
\author{Keith G. Ray}
\affiliation{Quantum Simulations Group, Lawrence Livermore National Laboratory, Livermore, CA 94550, USA}
\author{Chang-Eun Kim}
\affiliation{Quantum Simulations Group, Lawrence Livermore National Laboratory, Livermore, CA 94550, USA}
\author{Yuan Liu}
\affiliation{Center for Ultracold Atoms, Research Laboratory of Electronics, Massachusetts Institute of Technology, Cambridge, MA 02139, USA}
\author{Brenda M. Rubenstein}
\affiliation{Department of Chemistry, Brown University, Providence, RI 02912, USA}
\author{Vincenzo Lordi}  
\affiliation{Quantum Simulations Group, Lawrence Livermore National Laboratory, Livermore, CA 94550, USA}

\begin{abstract}

Ion traps are promising architectures for implementing scalable quantum computing, but they suffer from excessive ``anomalous'' heating that prevents their full potential from being realized. This heating, which is orders of magnitude larger than that expected from Johnson-Nyquist noise, results in ion motion that leads to decoherence and reduced fidelity in quantum logic gates. The exact origin of anomalous heating is an open question, but experiments point to adsorbates on trap electrodes as a likely source. Many different models of anomalous heating have been proposed, but these models have yet to pinpoint the atomistic origin of the experimentally-observed  $1/\omega$ electric field noise scaling observed in ion traps at frequencies between 0.1-10 MHz. In this work, we perform the first computational study of the ion trap electric field noise produced by the motions of multiple monolayers of adsorbates described by first principles potentials. In so doing, we show that correlated adsorbate motions play a definitive role in producing $1/\omega$ noise and identify candidate collective adsorbate motions, including translational and rotational motions of adsorbate patches and multilayer exchanges, that give rise to $1/\omega$ scaling at the MHz frequencies typically employed in ion traps. These results demonstrate that multi-adsorbate systems, even simple ones, can give rise to a set of activated motions that can produce the $1/\omega$ noise observed in ion traps and that collective, rather than individual, adsorbate motions are much more likely to give rise to low-frequency heating.
\end{abstract}

\maketitle


\section{Introduction}
\label{Introduction}

Despite their outstanding potential, fault-tolerant quantum computers capable of scaling to many qubits remain beyond our current grasp, in large part due to noise.\cite{Preskill_Quantum_2018} One particularly pernicious and ubiquitous form of noise that limits the coherence times of trapped ion,\cite{Wineland_Noise, Turchette_PRA_2000, Brownnutt_RMP,Brown2021} superconducting qubit, \cite{Wang_Schoelkopf_APL, Martinis_2009,Lee_PRL_2014} Rydberg atom,\cite{Saffman_2016} nitrogen-vacancy center-based,\cite{PhysRevLett.115.087602} and many other quantum architectures is surface noise. In ion traps -- in which cations that possess multiple, addressable hyperfine states are confined near metallic electrode surfaces -- surface-derived electric field noise has long been known to limit the fidelity of quantum gates by heating the very ion motions on which these gates depend.\cite{Brownnutt_RMP} The exact microscopic source of this electric field noise, however, remains far less understood. Although intuition would reflexively point to Johnson noise\cite{Johnson_PhysRev,Nyquist} -- the thermal noise generated by currents that run through the trap electrodes and circuitry -- as the dominant source of noise, a number of researchers have independently shown that a different, still unknown and therefore ``anomalous'' source of noise dominates the Johnson noise in ion traps.\cite{Hite_Surface_Science,Brown2021} The origin of this noise thus remains an open question in ion trap science whose resolution would dramatically improve the performance of ion trap-based quantum information processors,\cite{Haffner_Blatt,Benhelm2008,Bruzewicz_AppPhysRev} quantum sensors,\cite{McCormick2019,Degan_RMP,Baumgart_PRL,Leibfried1476} and clocks.\cite{Kozlov_RMP,Chou_PRL,Peik_PRL} 

Experiments directed at characterizing noise have revealed that the three key parameters that control the electric field noise, $S_{E}$, are the distance at which the ion is trapped above the electrode surface, $d$, the trapping frequency, $\omega$, and the trap temperature, $T$, such that 

\begin{equation}
    S_{E} \propto \omega^{-\alpha} d^{-\beta} f(T), 
\end{equation}
where $f(T)$ is a function of the temperature.\cite{Brownnutt_RMP} Although aspects of this functional form and its exponents are still a subject of debate,\cite{An_PRA_2019,Brownnutt_RMP} most experiments point to a frequency scaling involving an $\alpha$ of between 0 and 2 at MHz frequencies.\cite{Turchette_PRA_2000, Daniilidis_2011, Sedlacek_PRA_2018, Sedlacek_PRA_2018_2, Noel_PRA_2019,Labaziewicz_PRL_2008} This scaling is in line with the practical experience that most sources of noise have a ``$1/\omega$'' character with $\alpha=1$. Many of these same experiments also suggest a distance scaling of $d^{-4}$ with $\beta=4$,\cite{Turchette_PRA_2000, Sedlacek_PRA_2018, Sedlacek_PRA_2018_2, Deslauriers_PRL_2006,Daniilidis_2011, Hite_Pappas_2017,Labaziewicz_PRL_2008} which stands in stark contrast with Johnson noise, which scales as $d^{-2}$.  

Over the past decade, a number of different microscopic models\cite{Noel_PRA_2019,Kumph_2016,Teller_2021} have been advanced that aim to both reproduce and explain these scalings, which serve as useful constraints on the possible mechanisms that could give rise to anomalous heating. One of the earliest models proposed was the patch potential model,\cite{Turchette_PRA_2000} which espouses that local variations of electrode potentials can induce ion motion with the $d^{-4}$ noise scaling observed in experiments. The patch potential model, however, does not identify the source of the local fluctuations on which it is based. In light of experiments demonstrating that different treatments that remove surface adsorbates, including ion milling,\cite{Hite_PRL_2012,Daniilidis_PRB_2014,McKay_arXiv} plasma treatment,\cite{McConnell_PRA_2015} and laser cleaning,\cite{Allcock_2011} reduced trap noise by up to two orders of magnitude, it is logical to attribute this noise to dipole fluctuations caused by adsorbates bound to trap electrodes. 

Along these lines, two particularly compelling noise models are the adatom dipole and diffusion models. The adatom dipole model posits that atoms and molecules that adsorb onto electrode surfaces to form layers or patches develop induced dipole moments that locally lower the work function, and therefore the potential, of the metal.\cite{SafaviNaini_PRA_2011,Brownnutt_RMP} The dipoles may then be caused to fluctuate by phonon-induced transitions among different vibrational states of the adatom-surface potential, giving rise to the requisite fluctuation spectra. Initial investigations of this model that treated single adsorbates using plausible values for the their masses and forms for the their binding potentials illustrated that it can yield electric field fluctuation spectra that are of the same magnitude as those observed in experiments.\cite{SafaviNaini_PRA_2011,Safavi_Naini_PRA_2013} These spectra were flat as a function of frequency at low frequencies (the `white noise' regime), yet transitioned to decaying as $1/\omega^{2}$ at high frequencies with a brief $1/\omega$ regime in between. Such behavior is consistent with two-level fluctuator models prevalent in the signal processing literature.\cite{Dutta_RMP,Zanolla,Weissman_RMP} More recent work employing a first principles treatment of single hydrocarbon, water, and other adsorbate binding potentials and dipole moments similarly exhibited a narrow $1/\omega$ region in its spectra.\cite{Ray_2019} That work additionally included in-plane adsorbate vibrations, which produced the highest electric field noise and exhibited $1/\omega$ scaling at low frequencies orders of magnitude larger than the typical 1 MHz trap frequency. Interestingly, work released during the submission of this manuscript has suggested that phonons that correlate the motions of multiple adsorbates may also produce $1/\omega$ frequency scalings.\cite{Lloyd_arxiv} In the adatom diffusion model, dipole fluctuations are instead viewed as originating from the diffusion of an adataom or adsorbate across the surface of an electrode.\cite{Kim_PRA_2017} Noninteracting adatoms diffusing over a surface covered with patches with differing work functions may be shown to produce fluctuation spectra that scale as $1/\omega^{3/2}$.\cite{Hite_MRS_2013,Gesley_PRB_1985} 
Even though these models do not produce extended regions with $1/\omega$ frequency scalings in their current forms, the inclusion of realistic features such as adatom-adatom interactions and surface corrugation may lead to more complicated spectral features that have yet to be illuminated or understood. This thus raises the question of whether a more realistic treatment of adatom dipole fluctuations that incorporates diffusion and interparticle interactions may exhibit the so far \textit{theoretically} elusive $1/\omega$ scaling expected from experiments. 

In this work, we examine the frequency-dependence of the electric field fluctuation spectra produced by the dynamics of multiple monolayers of interacting methane adsorbates on gold, a commonly-used electrode material in ion traps. To study the adsorbate dynamics, we modeled them using classical Molecular Dynamics (MD) on  potential energy surfaces (PES) constructed based on first principles Density Functional Theory (DFT) calculations. From the MD trajectories, we obtained coverage-  and  temperature-dependent dipole-dipole fluctuation spectra using DFT-derived dipole moments. This enabled us to make a clear connection between the features of the fluctuation spectra at different frequencies and the underlying adsorbate motions that give rise to these features. 

In doing so, we demonstrate that correlated dipole fluctuations are overwhelmingly responsible for producing ``$1/\omega$'' noise and identify several key microscopic motions that naturally lead to a ``$1/\omega$'' frequency scaling in the MHz regime used in most ion traps. In particular, we find that correlated rotational and translational motions of adsorbates within clusters can give rise to ``$1/\omega$'' electric field noise at submonolayer coverages, while interlayer particle exchanges among the first two layers of adsorbates are the largest contributors to such noise at supermonolayer coverages. We furthermore show how many of the MHz features of the electric field spectra for these systems may be reproduced by a two-level fluctuator model with these motions as rare events. Even though the methane-gold system we study here is a model system that contains simplifications, our work provides clear evidence for the types of adsorbate motions that can give rise to anomalous heating and demonstrates the crucial role that collective, rather than individual, adsorbate motions assume in the noise generation process at experimental trap frequencies.

This paper is organized as follows. In Section \ref{Methods}, we begin by describing how we model the dipole-dipole fluctuation spectra of methane on a gold substrate using our combined DFT-MD approach. We next describe our results, including our key findings regarding the $1/\omega$-$1/\omega^{2}$ frequency-dependence of the dipole-dipole fluctuation spectra we obtain for varying adsorbate surface coverages and temperatures in Section \ref{Results}. In the same section, we additionally present our data directly linking specific adsorbate motions with features of the frequency spectra and demonstrate how our spectra can be reproduced using simple two-state models. Lastly, in Section \ref{Conclusions}, we place our findings that a realistic model of multilayer adsorbate dynamics can give rise to $1/\omega$ noise in the context of the ongoing search for the microscopic origins of anomalous heating, and discuss the limitations and natural extensions of our current model. Additional information regarding our simulations and interpretation of the data may be found in the Supplemental Information.

\section{Methods}
\label{Methods}

In this work, we model methane adsorbate dynamics on gold substrates at a range of temperatures and surface coverages by running classical MD trajectories on a DFT-derived substrate potential energy surface (PES). Based upon the trajectories obtained, we then compute adsorbate dipole-dipole correlation functions and Fourier transform them to acquire dipole-dipole fluctuation spectra. These spectra are subsequently analyzed for their frequency-dependent behavior, which can be used to predict ion trap noise.

\subsection{Calculation of Heating Rates}

As has been shown in previous work,\cite{Turchette_PRA_2000} the coupling of electric field fluctuations with the motion of trapped ions gives rise to a heating rate, $\dot{\hat{n}}$, that is given by: 
\begin{equation}
    \dot{\hat{n}} = \frac{q^2}{4m_I \hbar \omega_t} S_E(\omega_t).
    \label{eqn:noise}
\end{equation}
Here, $q$ is the charge of the ion, $m_I$ is the mass of the ion, $\omega_t$ is the frequency at which the ion is trapped (typically 0.1-10 MHz), and $\hbar$ is the reduced Planck's constant. $S_E(\omega)$ is the frequency spectrum of the electric field fluctuations. Patch potential models assert that electric field noise in traps stems from local potential fluctuations above the surfaces of the electrodes.\cite{Brownnutt_RMP, Dubessy_PRA_2009, Rossi_1992} These fluctuations can arise from regions of the electrodes with varying crystal orientations or adsorbate surface motion.\cite{} In the latter case, which is the focus of this investigation, fluctuations in the dipole moment of the adsorbates caused by electronic interactions with the electrode surface give rise to electric field noise. Specifically, for a conventional planar trap,\cite{Brownnutt_RMP} $S_E(\omega)$ may be obtained from out-of-plane adsorbate dipole fluctuations in a surface patch by:\cite{Ray_2019}

\begin{equation}
\label{eqn:S_E_out_of_plane}
    S_E(\omega) = \frac{3 \pi \sigma S_{\mu}(\omega)}{2(4 \pi \epsilon_0)^2 d^4} 
\end{equation} 
and from in-plane fluctuations by:

\begin{equation}
\label{eqn:S_E_in_plane}
    S_E(\omega) = \frac{\pi \sigma S_{\mu}(\omega)}{(4 \pi \epsilon_0)^2 d^4}.
\end{equation}
In Equations \eqref{eqn:S_E_out_of_plane} and \eqref{eqn:S_E_in_plane}, $\sigma$ is the average area density of the surface patches, $d$ is the ion-electrode distance, $\epsilon_0$ is the permittivity of free space, and $S_{\mu}(\omega)$ is the dipole-dipole fluctuation spectrum. Notably, these expressions for $S_E(\omega)$ reflect the $d^{-4}$ scaling that has often been observed experimentally in heating rates and is one key reason why we and others continue to examine this model. 

The dipole-dipole fluctuation spectrum of a surface patch can be calculated by taking the Fourier Transform of the dipole-dipole autocorrelation function:
\begin{equation}
\label{eqn:S_mu}
    S_{\mu}(\omega) = \int_{-\infty}^{+\infty} d\tau   e^{i \omega \tau} C_{\mu, \mu}(\tau), 
\end{equation}
where $C_{\mu, \mu}(\tau)$ represents the dipole-dipole autocorrelation function of the total patch dipole moment at time $\tau$. This autocorrelation function may be expressed as:

\begin{eqnarray}
\label{eqn:ACF}
C_{\mu, \mu}(\tau) &=&   \langle (\mu_{z}(\tau) - \langle \mu_{z} \rangle)(\mu_{z}(0) - \langle \mu_{z} \rangle) \rangle \nonumber \\
 &=&   \langle \Delta\mu_{z}(\tau)\Delta\mu_{z}(0)  \rangle  . 
\end{eqnarray}
In the above, $\mu_{z}(\tau)$ is the $z$-component of the total dipole moment of a patch at time $\tau$ and $\langle \mu_{z} \rangle$ is its equilibrium average.
Although a full treatment would use $\vec{\mu}$, we instead use $\mu_{z}$ unless otherwise specified. The values of the $x$ and $y$ components of our adsorbate dipoles are orders of magnitude smaller than the $z$ components. There are only a small handful of cases where the $x$ or $y$ components would affect spectra generated by $z$ components alone (see Supplemental Information).

In this work, we assume that $C_{\mu, \mu}(\tau)$ stems from how the dipole moments of $N$ different adsorbates change as they move across a surface with a spatially-varying binding potential. The quantity $\Delta\mu_{z}(\tau)$ represents a fluctuation of the electric dipole moment of the simulation cell, which is a sum over the individual adsorbate dipole fluctuations. This definition remains valid for Equations \eqref{eqn:S_E_out_of_plane} and \eqref{eqn:S_E_in_plane} as long as the simulation cell is small compared to the ion-electrode distance, $d$, and it captures the effect of adsorbate correlations on the noise as long as the simulation cell is larger than the correlation length.\cite{Brownnutt_RMP} The first condition is satisfied as the typical ion-electrode distance is 40 $\mathrm{\mu}$m and the simulation cell dimensions are less than 100 nm. The adsorbate-adsorbate correlation length is shown to be smaller than the simulation size in Section \ref{sec:correlatedspectra}. 

In this limit, we may express the surface patch dipole-dipole correlation function in terms of the adsorbate dipole fluctuations as:

\begin{eqnarray}
C_{\mu, \mu}(\tau) &=&   \Big \langle \sum_{i}^N  \Delta\mu_{i,z}(\tau)  \sum_{j}^N  \Delta\mu_{j,z}(0)  \Big\rangle \label{eqn:decompose} \\
&=&   \Big \langle \sum_{\substack{i,j \\ i= j}}^{N,N}  \Delta\mu_{i,z}(\tau)  \Delta\mu_{j,z}(0)  \Big\rangle \nonumber \\ 
&+& \Big \langle  \sum_{\substack{i,j \\ i\neq j}}^{N,N}  \Delta\mu_{i,z}(\tau)  \Delta\mu_{j,z}(0)  \Big\rangle, \nonumber
\end{eqnarray}
where $\Delta\mu_{i,z}(\tau)$ represents the fluctuation of the $z$ component of the electric dipole moment of adsorbate $i$ at time $\tau$. In the last expression, we have divided the sum into a sum over dipole-dipole correlation functions of the individual ($i = j$) adsorbates and a sum over those correlations between distinct ($i \neq j$) adsorbates. If there were no correlation between different adsorbates, the second term would vanish. We calculate both of these sums and demonstrate the importance of dipole-dipole correlations between different adsorbates in Section \ref{sec:correlatedspectra}.

In practice, we evaluate Equation \eqref{eqn:ACF} by averaging over all of the individual adsorbate dipole-dipole autocorrelation functions taken across the entire simulation runtime, $\mathcal{T}$:
\begin{multline}
\quad
 C_{\mu,\mu}(\tau) =\\ 
  \frac{1}{(\mathcal{T}-\tau)} \sum_{k=1}^{\mathcal{T}-\tau}  \left[ \sum_{i}^N  \Delta\mu_{i,z}(\tau + t_k)  \sum_{j}^N  \Delta\mu_{j,z}(t_k) \right],  
        \label{eqn:cfxn_biased}
\end{multline}
where $\mathcal{T} - \tau$ denotes the number of timesteps taken between these two times. 

\subsection{Generation of the Electrode Potential Energy Surface \label{Gen_PES}}
In order to calculate the fluctuations of the dipole moments required by Equation \eqref{eqn:cfxn_biased}, a PES of \ch{CH4} physisorbed onto Au(111) was interpolated from DFT \ch{CH4}-Au(111) binding energies taken at a representative set of positions on the Au(111) surface. In all of our simulations, the methane adsorbates were treated as point particles, a reasonable simplification given their spherical character, and we did not consider the effects of phonons.

\begin{figure*}
\centering
\includegraphics[width=\linewidth]{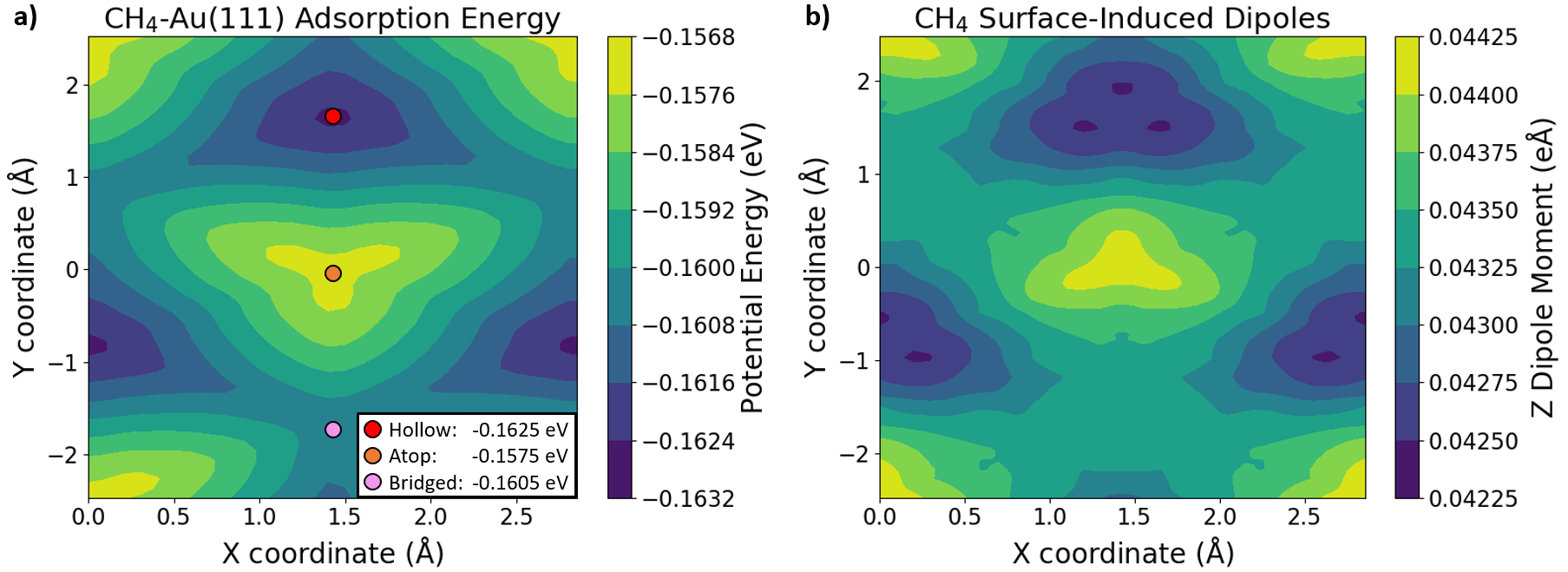} 
\caption{{\bf (a)} Visualization of the \ch{CH4}-Au(111) potential energy surface generated from DFT calculations spanning one Au(111) unit cell at an adsorbate-surface distance of 3.16 \r{A}, which is the minimum-energy, out-of-plane distance. The red, orange, and pink circles designate the positions of the Au(111) hollow, atop, and bridged sites, respectively. {\bf (b)} Visualization of the magnitudes of the surface-induced methane dipole moments as a function of methane position in the same Au(111) cell and at the same height as on the left.} 
\label{fig:Potential_Figure}
\end{figure*}

The DFT calculations were performed with VASP\cite{Kresse_PRB} using the vdW-DF-cx functional \cite{PhysRevB.89.035412} at a single k-point and a plane-wave cutoff of 600 eV. The resulting potential energy points were interpolated across a periodic surface. The interpolated PES employed has dimensions of 2.856 \r{A}, 4.947 \r{A}, and 29.0 \r{A} in the $x$, $y$, and $z$ directions, respectively. Figure \ref{fig:Potential_Figure}a depicts this PES in the $(x,y)$-plane for one Au(111) unit cell at the minimum-energy distance from the carbon of the adsorbate to the surface  of 3.16 \r{A}. The simulation cell contained 72 such unit cells (twelve in the $x$ direction and six in the $y$ direction). This unit cell tiling was chosen so as to make the dimensions of the overall simulation cell roughly the same in both the $x$ and $y$ directions to reduce any spurious adsorbate ordering effects from anisotropic boundary conditions. As can be seen in Figure \ref{fig:Potential_Figure}a, the hollow site marks the lowest-energy binding position on this surface (-0.1625 eV), while the atop site marks the highest-energy position (-0.1575 eV). Figure S3 
projects the PES along the $z$ direction. Additional simulation parameters may be found in Table \ref{tab:simulation_parameters}, and additional binding energies at different positions on the surface may be found in Table SI. 

In addition to the surface-adsorbate interaction, each adsorbate was modeled as interacting with its neighbors through the methane-methane Optimized Potentials for Liquid Simulations (OPLS) potential\cite{Jorgensen_JACS} such that:
\begin{equation}
V_{adsorbate} = V_{surface} + V_{interparticle}, 
\end{equation}
where 
\begin{equation}
V_{interparticle} = 4 \epsilon_{CH_4} \left[ \left( \frac{\sigma_{CH_4}}{r_{ij}} \right)^{12} -  \left( \frac{\sigma_{CH_4}}{r_{ij}}\right)^{6} \right]. 
\end{equation}
In this equation, $\epsilon_{CH_4}$ denotes the OPLS \ch{CH4}-\ch{CH4} interaction energy, while $\sigma_{CH_4}$ denotes the \ch{CH4}-\ch{CH4} interaction radius. $r_{ij}$ represents the interparticle distance between the centers of mass of methanes $i$ and $j$.  Based upon the OPLS interaction radius, the methanes would prefer to be spaced 4.19 \r{A} from one another. As we will see, this spacing favored by the OPLS potential competes with the particle spacing favored by the surface potential to produce different particle surface configurations at different temperatures, which in turn give rise to the different types of particle motions that are reflected in the dipole spectra. 

\subsection{Molecular Dynamics Using the Atomic Simulation Environment} \label{sec:md}

The Atomic Simulation Environment (ASE) was used to perform the MD simulations with the interpolated \ch{CH4}-Au(111) and OPLS potentials. Running MD simulations based upon this interpolated landscape is much more computationally efficient than running fully \textit{ab initio} simulations and is therefore key to reaching the timescales required to observe events happening in the 0.1-10 MHz frequency range at which ions are trapped.

ASE is an open-source software platform for atomistic simulations.\cite{asepaper} It is also Python-based, which makes it possible to seamlessly incorporate the Python-interpolated PES directly into the MD setup. The MD scripts used can be found in this paper's Github repository.\cite{Github}

NVT MD simulations were performed using Langevin dynamics with a friction coefficient of 0.01 atomic units (a.u., where 1 a.u. $= 4.13\times 10^{16}$ s$^{-1}$). Particle forces and energies were updated every 5 fs, and positions were recorded for analysis every 1 ps to ensure that configurations were not artificially correlated. These parameters were verified to be computationally efficient while providing comparable system equilibration and dynamics to those obtained using smaller friction coefficients and step sizes (see Figures S1 and S2).
In order to isolate collective motions of interest from thermostat-induced cluster translations and drifts, the thermostat was configured to fix the center-of-mass position and zero the center-of-mass momentum. Simulation runtimes ranged from 6 to 19 $\mathrm{\mu}$s for different sets of coverages (see Supplementary Info). The MD cell employed periodic boundary conditions in the $x$ and $y$ directions, but not in the $z$ direction.

\begin{table}
    \centering
    \renewcommand\arraystretch{1.25}
    \renewcommand{\tabcolsep}{8pt}
    \begin{tabular}{c|c}
        \hline
        \ch{CH4}-Au(111) Potential & DFT, vdW-DF-cx \\
        \ch{CH4}-Au(111) min. energy distance & 3.16 \r{A} \\
        \ch{CH4}-Au(111) min. energy & -0.1625 eV 
        \\
        OPLS min. energy distance & 4.19 \r{A} \\
        $\sigma_{\ch{CH4}}$ for OPLS & 3.73 \r{A} \\
        $\epsilon_{\ch{CH4}}$ for OPLS & 12.75 meV \\ 
        MD Environment & ASE \\
        Thermostat & Langevin \\
        Friction Coefficient & 0.01 a.u. \\
        MD Force Update Step & 5 fs \\
        MD Position Recording Step & 1 ps \\
        \hline
    \end{tabular}
    \caption{Simulation details and parameters.}
    \label{tab:simulation_parameters}
\end{table}

\subsection{Calculation of Dipole Moments}
The electric dipole moments of the methanes adsorbed over a range of positions on the gold surface were also computed with DFT using VASP. In particular, we integrated the charge density $\times$ position, $\rho(r)\times r$, over a 40$\times$33.6$\times$29.1 \r{A}$^{3}$ simulation cell containing an Au slab ``island'' with a finite extent in the in-plane directions and a single \ch{CH4} molecule on top. A finite slab is required so that, if the adsorbate is moved within one Au surface unit, charge is not induced to move across the periodic boundaries and the integral remains well-defined. The slab island consists of a 4$\times$4 arrangement of Au surface primitive cells. To map the surface-induced dipole moments as depicted in Figure \ref{fig:Potential_Figure}b, we calculated the dipole moments for \ch{CH4} at the surface positions illustrated in one surface unit cell located in the center of this slab and at varying heights. Electric dipole corrections were utilized in VASP and a small linear ramp in the dipole moment was subtracted to remove remaining finite-size effects and ensure that the dipole moment is periodic with respect to the surface in-plane periodicity. As in Section \ref{Gen_PES}, these calculations were also performed using the vdW-DF-cx functional at a single k-point and with a 600 eV plane-wave cutoff.

The resulting dipole moments were interpolated using an interpolation script to yield the $x$, $y$, and $z$ components of the dipole vector as a function of adsorbate position. The values of the $x$ and $y$ components of the dipole moments are negligible compared to the $z$ component values across the entire periodic cell, and their fluctuations do not produce much noise except in a handful of cases. Thus, we chose to make use of the $z$ components of the dipole moments for the rest of this study (see the Supplemental Information for supporting data).  A visualization of the $z$ dipole moment surface at the minimum-energy out-of-plane surface-adsorbate distance can be found in Figure \ref{fig:Potential_Figure}b. Comparing the dipole moment surface with the PES, it can be seen that the largest induced dipole moments occur at the locations with the largest potential energies, and vice versa. The dipole interpolation script that produced this surface is provided in this paper's Github repository.\cite{Github}

\subsection{Dipole-Dipole Fluctuation Spectra}

Dipole-dipole autocorrelation functions were computed using Equation \eqref{eqn:cfxn_biased} based upon how the adsorbate dipole moments change in time as they traverse the potential surface. As we modeled systems with varying surface coverages (see Table \ref{tab:run_table}), the exact value of $N$ used in this equation depended upon the coverage studied. The first 1,000 timesteps (spanning 1 ns) were omitted to allow for equilibration. Dipole-dipole fluctuation spectra were then calculated by applying a Discrete Fast Fourier Transform\cite{1989ApJ} to these correlation functions (per Equation \eqref{eqn:S_mu}). A standard smoothing technique, Blackman smoothing,\cite{BMwindow} was applied before plotting.

\subsection{Two-Level Fluctuators as a Source of $1/\omega^2$ and $1/\omega$ Noise}
\label{sec:TLF_overview}

As will be discussed further below, our electric field fluctuation spectra exhibit several key frequency scalings. Many of these frequency scalings have well-known connections to certain types of fluctuations in the signal processing literature that also arise in our simulations. For instance, a quantity that fluctuates randomly about a single average value with no correlation in time leads to featureless (white) noise\cite{shynk_2012} in the frequency domain. On the other hand, a quantity that transitions between two discrete values produces a region of $1/\omega^2$ frequency scaling; systems that fluctuate in this manner are known as two-level fluctuators\cite{Dutta_RMP} (TLFs) or random telegraph signals (RTS).\cite{Zanolla} 

More specifically, a quantity that switches between states 0 and 1 differing in value by $\Delta I$ and that has mean state residence times of $\tau_0$ and $\tau_1$ will yield noise spectra of the following Lorentzian form:\cite{Machlup}
\begin{equation}
    S(\omega) = \frac{4(\Delta I)^2}{(\tau_0 + \tau_1)[(1/\tau_0 + 1/\tau_1)^2 + \omega^2]}.
\label{eqn:TLF}
\end{equation}
When $\tau_0 = \tau_1 = \tau$, this expression simplifies to:

\begin{equation}
    S(\omega) = \frac{4(\Delta I)^2\tau}{1 + \tau^2\omega^2}.
\label{eqn:TLF_simplified}
\end{equation}

Equations \eqref{eqn:TLF} and \eqref{eqn:TLF_simplified} produce white noise at low frequencies followed by a rounding-off to a $1/\omega^2$ descent at frequencies greater than a cutoff frequency, given by $\omega_{c} = (1/\tau_0 + 1/\tau_1)$. 
Figure S16 shows how TLF noise spectra vary with $\tau_0$ if $\tau_1$ is held constant. When $\tau_1$ is constant, the TLF noise reaches a maximum when $\tau_0=\tau_1$. When $\tau_0>\tau_1$, the TLF noise decreases while $\omega_{c}$ remains mostly unchanged; when $\tau_0<\tau_1$, the TLF noise decreases and $\omega_{c}$ increases, effectively shifting the frequency curve to the right.\cite{Zanolla} In general, decreasing $\tau_0$ and $\tau_1$ will shift the $1/\omega^2$ region to higher frequencies.

In addition to a $1/\omega^2$ scaling, TLFs can give rise to $1/\omega$ regions in several ways. Every individual TLF has sections of $1/\omega$ and near-$1/\omega$ scaling in the rounding-off region between the white noise and $1/\omega^2$ regimes.
Moreover, an ensemble of TLFs with different transition frequencies can produce an aggregate $1/\omega$ region over several decades of frequencies;\cite{Zanolla} fluctuators with more than two levels can also combine together to produce $1/\omega$ regions in a similar manner.


\section{Results and Discussion}
\label{Results}

\subsection{Adsorbate Surface Geometries}
\label{sec:ads_surface_geometries}

\begin{figure*}[!htbp]
\centering
\includegraphics[width=\linewidth]{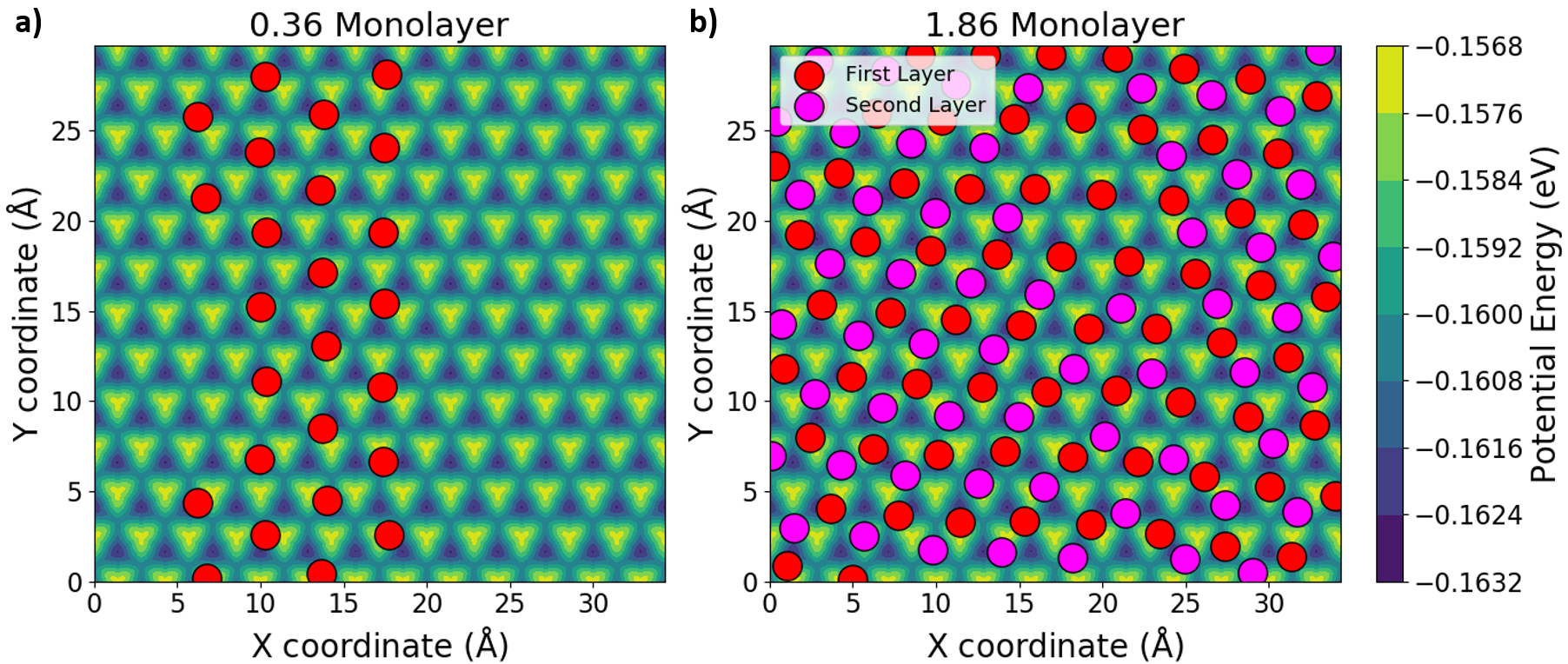}
\caption{Representative snapshots of the positions of the adsorbates on the Au(111) surface at 20 K for {\bf (a)} 0.36 ML and {\bf (b)} 1.86 ML coverages. Note that the intermolecular potential is stronger than the in-plane Au(111)-\ch{CH4} potential barriers when the particles are near the surface. Because the adsorbates cannot simultaneously reside in the differing minima of their intermolecular and surface-adsorbate potentials, they do not all occupy lowest energy hollow sites on the surface; instead, they organize into a more complicated, but repeating pattern. First layer particles can have a maximum of six first layer and three second layer neighbors.}
\label{fig:coverage_panel}
\end{figure*}

We begin by analyzing the geometries that the adsorbates assume at the surface coverages studied. Because of the competing influences of the interparticle and surface-adsorbate potentials, the geometries are not straightforward to predict. Here, we visualize and discuss several emblematic coverages.

Within the simulation box employed in this work, we find that 69 adsorbates produce the full monolayer (ML) surface coverage with the lowest energy per adsorbate (see Figure S10).
We thus define our runs in terms of $(N/69)$ ML coverages. In Figure \ref{fig:coverage_panel}, for example, the 0.36 ML coverage on the left has 25 particles, whereas the 1.86 ML coverage on the right has 128 particles. As a guide to our subsequent discussion, we list all of the surface coverages and temperatures at which our simulations were performed in Table \ref{tab:run_table}.

\begin{table}
    \centering
    \renewcommand\arraystretch{1.25}
    \renewcommand{\tabcolsep}{4pt}
    \begin{tabular}{c|c}
        {\bf Type} & {\bf Coverage ($N_{\ch{CH4}}$)}\\
        \hline
        {\bf Submonolayer} & 0.36 ML (25)\\
        {\bf Submonolayer} & 0.52 ML (36)\\
        {\bf Submonolayer} & 0.71 ML (49)\\
        {\bf Monolayer} & 1.0 ML (69)\\
        {\bf Supermonolayer} & 1.16 ML (80)\\
        {\bf Supermonolayer} & 1.42 ML (98)\\
        {\bf Supermonolayer} & 1.86 ML (128)\\
    \end{tabular}
    
    \begin{tabular}{cc}
         & \\
    \end{tabular}
    
    \begin{tabular}{c|c}
        \hline
        {\bf Temp. $<$ 70 K} & 20, 25, 30, 35, 40, 45, 50, 55, 60 \\
        {\bf Temp. $\geq$ 70 K} & 70, 80, 90, 100, 110, 125, 150, 184, 220 \\
        \hline
    \end{tabular}
    \caption{Table of coverages and temperatures at which simulations were performed in this work.}
    \label{tab:run_table}
\end{table}
Figure \ref{fig:coverage_panel}a depicts a representative surface geometry of the 0.36 ML trajectory at 20 K. At this coverage, the methanes cluster together and simultaneously try to minimize their interparticle energies while residing in the hollow-site minima of the Au(111) surface. Notably, the methanes cannot reside at every available hollow site because such sites are separated by only 2.856 \r{A}, which is significantly less than $\sigma_{\ch{CH4}}$=3.73 \r{A}, the zero-energy OPLS distance. Since the OPLS well depth is greater than any surface site energy differences (see Tables \ref{tab:simulation_parameters} and SI),
the methanes occasionally settle for bridged or even atop sites in order to minimize interparticle repulsions. In so doing, they maintain a \ch{CH4}-\ch{CH4} distance of $>$3.73 \r{A}, close to the interparticle potential minimum-energy distance of 4.19 \r{A}. 
Habitually, the methanes lying near atop sites get pulled toward hollow sites and push neighboring hollow-site methanes into less favorable sites, and so on and so forth throughout the clusters. This in-plane rattling occurs often -- within 1-2 \r{A} and at much higher frequencies than trapped ion frequencies -- but does not disrupt the cohesion of the clusters at low temperatures. The methanes can also become perturbed enough to undergo more significant individual and collective movements, and while these significantly affect the noise spectra (see Section \ref{sec:submonolayer}), they do not prevent the surface clusters from remaining intact and mostly stationary at low temperatures. At higher temperatures, the particles have plenty of thermal energy to leave both the surface and \ch{CH4}-\ch{CH4} interaction wells, resulting in a more fluid and less-defined surface geometry.

Although most previous works have focused on adsorbate dynamics at submonolayer coverages assuming that diffusion across the electrode surface is the dominant source of noise,\cite{Kim_PRA_2017} here we additionally examine supermonolayer adsorbate dynamics. Studying multiple monolayers is more true to experimental reality, as several experiments have shown multiple hydrocarbon monolayers to be present on untreated gold traps.\cite{Hite_PRL_2012, Kim_PRA_2017} Figure \ref{fig:coverage_panel}b depicts a representative 1.86 ML bilayer coverage at 20 K. Not surprisingly, the second layer stacks on top of the first such that each second layer particle is roughly equidistant (in the $(x,y)$ plane) from its three nearest first layer neighbors, while typically staying at least 3.4-3.6 \r{A} above the first layer (6.56-6.76 \r{A} above the surface) in the $z$ direction. This leads to clustering in the second layer, similar to how clusters formed on the surface at submonolayer coverages.

Second layer clusters are affected by interparticle interactions from both first and second layer particles, but second layer particles are not as strongly affected by the surface potential. In fact, at typical second layer distances from the surface, the magnitude of the adsorbate-surface potential is comparable to the minimum OPLS energy (see Figure S3).
Beginning around 70 K, the second layer clusters become less coherent as interparticle interactions become less restrictive to particle motion. This leads to more fluid second layer particles, both parallel and perpendicular to the surface, some of which begin to stochastically exchange layers. Such exchanges can be reciprocal or unidirectional; in the latter case, the population of the surface changes. Such exchanges can be reciprocal or unidirectional; in the latter case, the population of the surface changes. Although 69-adsorbate surfaces are slightly favored in monolayer conditions, other surface populations can manifest in supermonolayer coverages, including 68- and 70-adsorbate surfaces. The switching between different surface populations will become important in our interpretation of the dipole-dipole fluctuation spectra.

Due to the dwindling influence of the surface-electrode potential at further distances from the electrode surface, the propensity to form a cohesive third layer is significantly reduced. In addition, the dwindling electrode-adsorbate interaction strength makes it more difficult to keep weakly-held second layer particles from getting pushed up or drifting to higher surface heights as the second layer gets more crowded. For these reasons, we limit our focus to a maximum coverage of 1.86 ML. More visualizations of and information about different coverages can be found in the Supplementary Information.

\subsection{Effects of Correlated Adsorbate Motion on Fluctuation Spectra}
\label{sec:correlatedspectra} 

\begin{figure}[!htbp]    \centering
    \includegraphics[width=\linewidth]{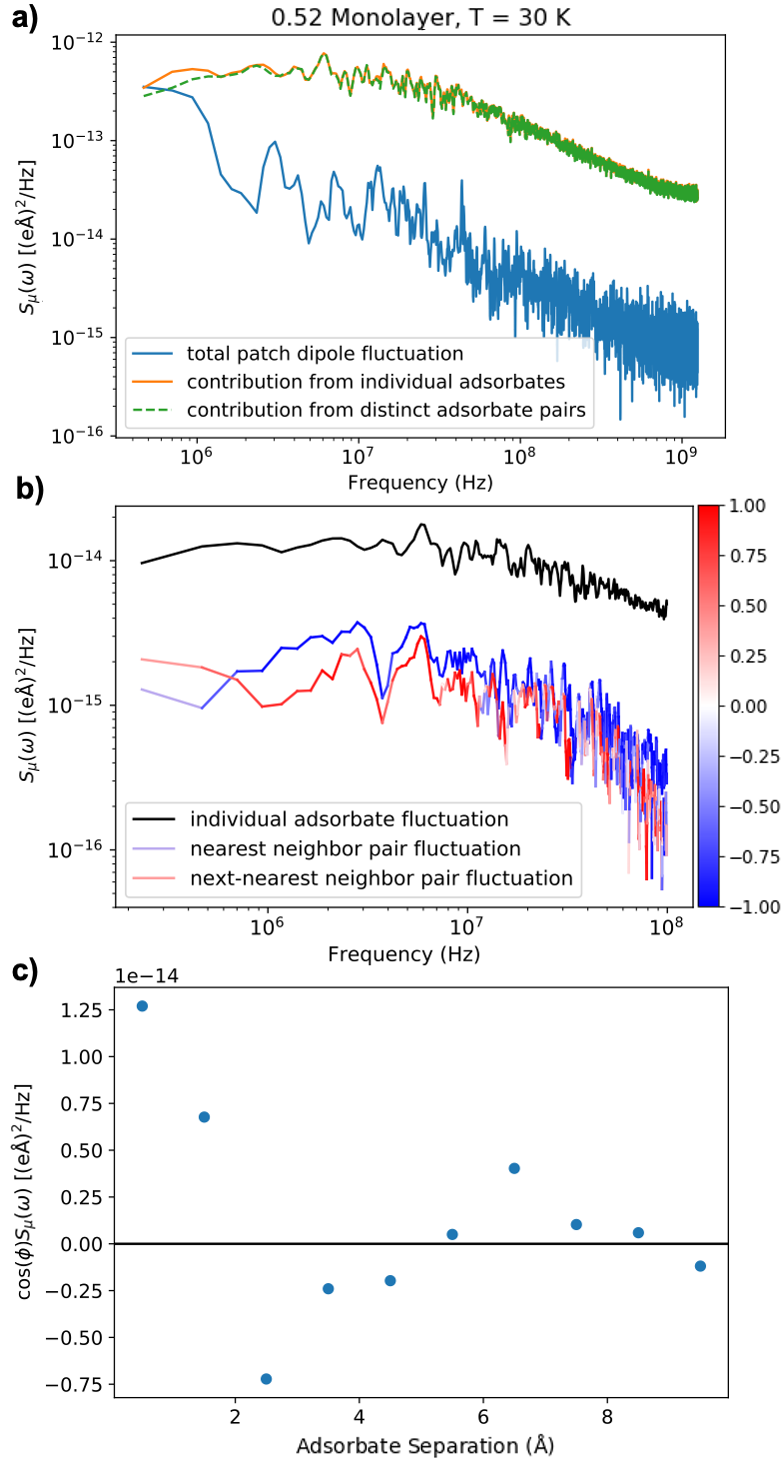}
    \caption{(a) Patch total dipole fluctuation spectrum, ($\mathcal{F}(\langle \sum  \Delta\mu_{i,z}(\tau) 
    \\ \Delta\mu_{j,z}(0)  \rangle)(\omega) $), in blue, and that spectrum decomposed into individual, orange ($i=j$), and distinct, dashed green ($i\neq j$) adsorbate contributions, which significantly cancel each other. (b) Dipole-dipole fluctuation spectrum of an individual adsorbate, black line, and the spectra of pairs of adsorbates separated by less than 6.0 \AA, mostly blue, and between 6.0 \AA \ to 9.0 \AA, mostly red, corresponding to surface adsorbate nearest neighbors and next-nearest neighbors. The colors of the nearest-neighbor and next-nearest neighbor pair spectra are indicative of their relative phase compared to the individual adsorbate fluctuations, with $\cos(\phi) = 1  (-1)$ being dark red (blue). (c)  $\cos(\phi)\mathcal{F}(\langle   \Delta\mu_{i,z}(\tau)  \Delta\mu_{j,z}(0)  \rangle)(\omega=\text{3.5 MHz})$ vs. distance between adsorbates $i$ and $j$. $\phi$ is the phase difference between the distinct $i/j$ adsorbate spectrum and the individual adsorbate spectrum at that frequency.}
    \label{fig:c_effect}
\end{figure}

To demonstrate the role that dipole-dipole correlations between distinct surface adsorbates play in the generation of the electric field noise affecting trapped ions, we decompose the total simulation patch dipole-dipole correlation spectrum into contributions from individual adsorbates and from distinct adsorbate pairs. This decomposition is expressed in Equation \eqref{eqn:decompose} and presented in Figure \ref{fig:c_effect} for the case of a 0.52 ML coverage of adsorbed \ch{CH4} at 30 K. The trends shown are representative of those also observed for different coverages and temperatures. We notice in Figure \ref{fig:c_effect}a that, while the individual adsorbate dipole-dipole spectrum and the distinct adsorbate dipole-dipole spectrum both exhibit roughly a white-noise spectrum up to 10 MHz, the total dipole-dipole correlation spectrum scales as $1/\omega$ beyond 10 MHz. Therefore, capturing dipole-dipole correlations between distinct adsorbates is essential for reproducing the $1/\omega$ electric field noise measured experimentally at frequencies of a few MHz.

Interestingly, the magnitude of the total dipole-dipole spectrum is smaller than that of either the individual or distinct pair adsorbate contributions. This occurs because the pair adsorbate dipole spectrum is close to the individual adsorbate dipole spectrum in magnitude, but is roughly $\pi$ phase shifted and thus partially cancels the noise from the individual adsorbate dipole spectrum. In Figure \ref{fig:c_effect}b, we investigate this behavior in more detail by plotting the average spectrum of an individual adsorbate along with the dipole spectra between adsorbates that are nearest and next-nearest neighbors, corresponding to adsorbate separations of less than 6.0 \AA \ and 6.0-9.0 \AA, respectively. Positions were averaged over a window of 2000 timesteps prior to spectral calculation. The colors of the plotted pair spectra indicate the phase as $\cos(\phi)$, with dark blue signifying $\cos(\phi) = -1$, $\phi = \pi$ and dark red indicating $\cos(\phi) =1$, $\phi = 0$. Both the magnitude of the spectrum and its color are indicative of the correlation with the individual adsorbate dipole fluctuation. Nearest-neighbor dipoles are slightly anticorrelated on average, indicated by the dark blue color and a magnitude that is smaller than that of the individual adsorbate spectrum, while next-nearest neighbors have an even smaller correlation that is positive on average. While the magnitude of the dipole-dipole spectrum of an average nearest-neighbor pair is less than that of the individual adsorbate spectrum, there are 5.3 nearest neighbors on average for this coverage at this temperature, so the sum of all distinct pairs produces the dashed curve in Figure \ref{fig:c_effect}a.

In Figure \ref{fig:c_effect}c, we plot $\cos(\phi)\mathcal{F}(\langle   \Delta\mu_{i,z}(\tau)  \Delta\mu_{j,z}(0)  \rangle)(\omega=\text{3.5 MHz}) $  vs. the distance between adsorbates $i$ and $j$, where $\phi$ is the phase difference between the distinct $i/j$ adsorbate spectrum and the individual adsorbate spectrum at that frequency and separation. From this plot, we can see how the magnitude and phase of the pair dipole-dipole fluctuation spectra evolve with separation.  As above, $\cos(\phi) = -1$ signifies anticorrelated adsorbate dipole fluctuations and  $\cos(\phi) = 1$ signifies correlated adsorbate dipole fluctuations, and adsorbate positions were averaged over a window of 2000 timesteps. In Figure \ref{fig:c_effect}c, we again see the anticorrelation of the nearest-neighbor adsorbates, which is strongest at 2.5 \AA \ (see also the blue line in Figure \ref{fig:c_effect}b). We note, however, that most nearest-neighbor adsorbates are close to 4.5 \AA \ apart, a distance at which the anticorrelation is weaker than at 2.5 \AA. We can quantify the correlation between adsorbates as close as an angstrom or two apart because we average over many adsorbate configurations and a few such rare configurations arise given enough sampling. The correlation of the next-nearest neighbors, first captured by the red line in Figure \ref{fig:c_effect}b, is positive and most next-nearest neighbor adsorbates are around 8 \AA \ apart. We estimate the correlation length to be 5.7 \AA, much smaller than our simulation cell.

\subsection{Submonolayer Dynamics and Spectra}
\label{sec:submonolayer} 

\begin{figure*}
    \centering
    \includegraphics[width=\linewidth]{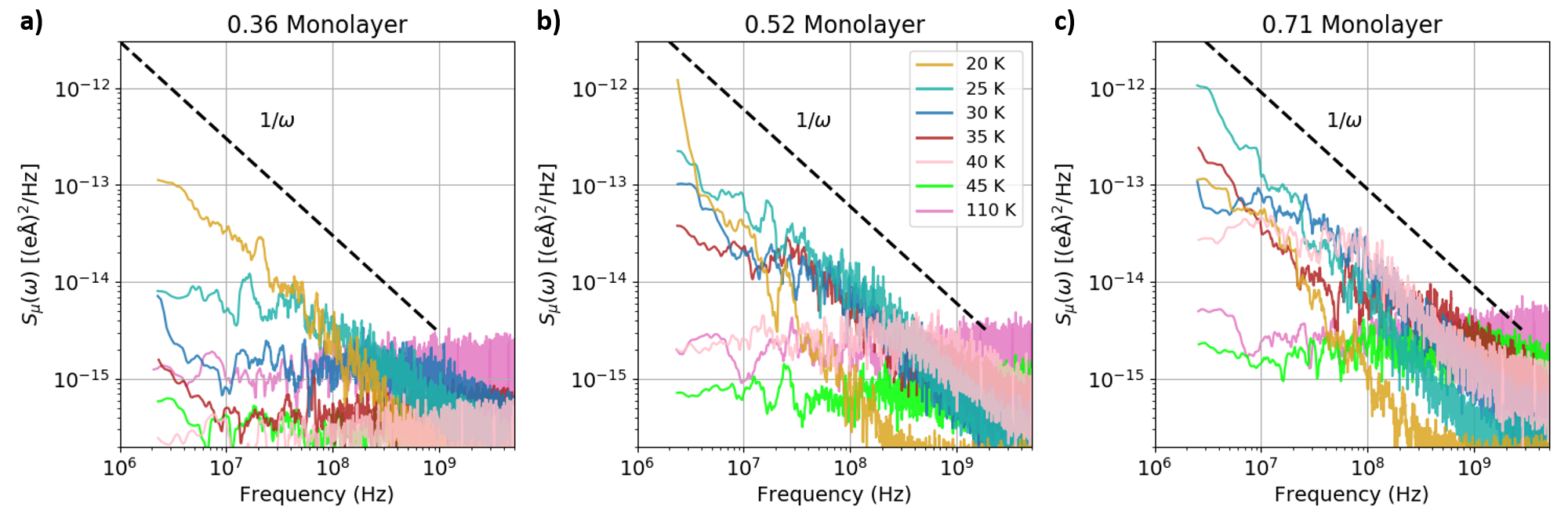}
    \caption{$S_{\mu}(\omega)$ spectra obtained from dipole moment time series data of the submonolayer coverage runs. The spectra were smoothed over a window of 12 timesteps.}
    \label{fig:submono_spectra}
\end{figure*}

Here, we examine the spectra produced by adsorbate dynamics at submonolayer coverages, as depicted in Figure \ref{fig:submono_spectra}. The spectra notably contain regions that scale as $1/\omega$ between 10$^6$ and 10$^9$ Hz, which overlaps with the range of typical  trap frequencies of 0.1-10 MHz.

The spectra can be divided into two general categories based on temperature. For moderate to high temperatures ($>$ 50 K), the noise is independent of frequency (white noise) and its magnitude increases with temperature. For low temperatures (20-40 K), the \textit{high-frequency} noise is also flat and greater at higher temperatures, but the \textit{low-frequency} noise behaves quite differently. At these temperatures, the noise levels increase with decreasing frequency with a mixture of $1/\omega$ and $1/\omega^2$ scalings before eventually flattening to a second white noise region. Moreover, increasing the temperature leads to a decrease in the low-frequency noise ceiling and an increase in the cutoff frequency at which the ceiling is reached; this is similar to the behavior of TLF-derived systems, as discussed in Section \ref{sec:TLF_overview}.

\begin{figure*}
    \centering
    \includegraphics[width=\linewidth]{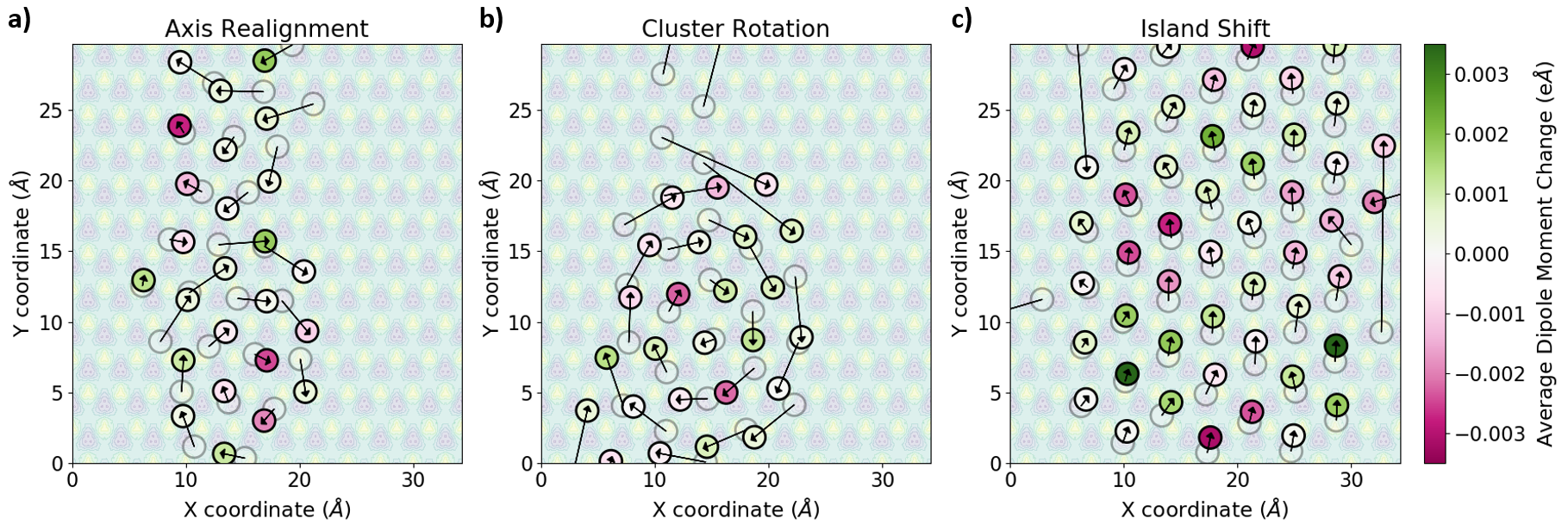}
    \caption{Selected collective motions observed in submonolayer trajectories. Arrows point from each particle's initial position (represented by the dimmed circles) to its final position. The Au(111) induced dipole surface (from Figure \ref{fig:Potential_Figure}b) is faintly shown beneath the particles. The Axis Realignment {\bf (a)}, in which the particles reorient around a new axis, and Cluster Rotation {\bf (b)} examples are taken from the 0.36 ML trajectory at 20 K, while the Island Shift {\bf (c)} example is from the 0.71 ML trajectory at 30 K. Such motions can take on the order of 10's or the low 100's of ps to occur.}
    \label{fig:collective_motions_panel}
\end{figure*}

\begin{figure}[h] 
\centering
    \includegraphics[width=\linewidth]{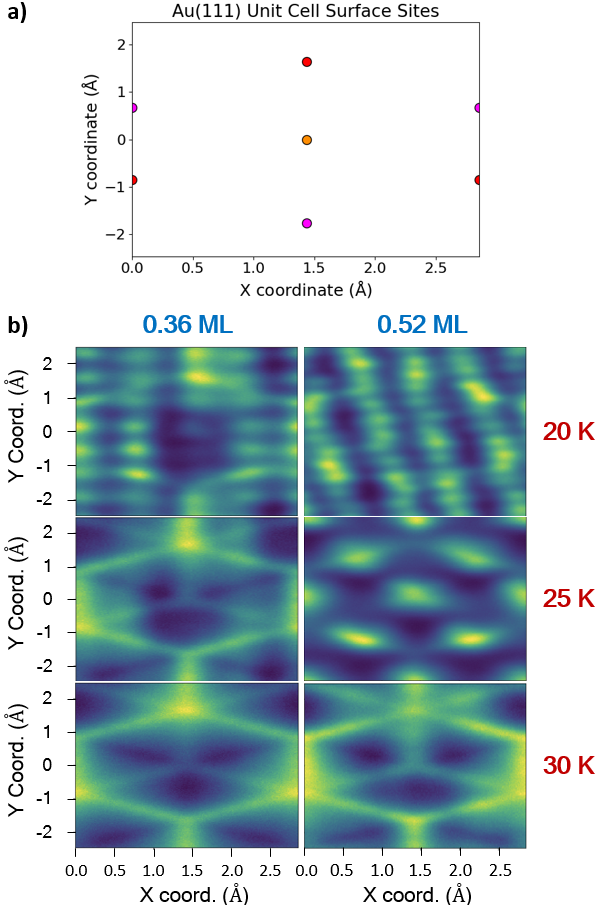}
    \caption{Two-dimensional normalized histograms of the relative positions of \ch{CH4} adsorbates on the Au(111) unit cell surface for selected submonolayer coverages. {\bf (a)} Surface facets (for reference). {\bf (b)} Unit cell adsorbate position histograms. At lower temperatures, a larger proportion of the particles reside near less favorable atop and bridged sites due to the strength of the adsorbate clustering. As the temperature increases, the adsorbate clusters become more fluid, and individual particles are able to migrate along the surface more often and end up residing in more favorable sites.
    }
    \label{fig:monolayer_cov}
\end{figure}

Each of the different aspects of the low temperature spectra can be matched to different types of movements. The high-frequency white noise stems from ubiquitous in-place adsorbate rattling of $<$2 \r{A} -- both in-plane and out-of-plane -- which causes random and uncorrelated dipole fluctuations. The low-frequency noise comes from movements -- both individual and collective (see Figure \ref{fig:collective_motions_panel}) -- that are sufficiently rare and substantial to give rise to different distinct dipole regimes between which the system fluctuates. Figure S17
illustrates this point by comparing the featureless spectrum of a particle with effectively no average dipole variation with the $1/\omega^2$-containing spectrum of a particle that transitioned between two distinct dipole regimes.

At low temperatures, the source -- and rarity -- of the motions responsible for low-frequency noise can be traced back to the cohesive strength of the adsorbate clusters. As Figure \ref{fig:monolayer_cov} shows, these clusters grow more cohesive with decreasing temperature and increasing $N$, and, at their strongest, are so cohesive that adsorbates routinely inhabit less-favorable surface facets in order to minimize their collective \ch{CH4}-\ch{CH4} interactions. In these clusters, each adsorbate remains in a distinct location on the surface: although the adsorbates frequently rattle in the $x,y$, and $z$ directions, their average surface positions do not change. Crucially, this also means that their average dipole moments do not change (although the continual rattling makes their dipole values very noisy as a function of time). However, when perturbed enough, the clusters can undergo reconfigurations. This can be caused by individual particle motions -- such as edge hopping -- or, more often, correlated multi-particle motions -- such as collective island shifts, internal ripples, and cluster rotations (Figure \ref{fig:collective_motions_panel}). Either way, these reconfigurations change the global surface positions -- and, with them, the average dipole moments -- of one or more adsorbates; dipole changes can result from changes in the surface sites or $z$ heights the different adsorbates occupy because of these motions. Such configuration-altering moves are rare at low temperatures since the \ch{CH4}-\ch{CH4} attraction is high and the system's kinetic energy is low (lower than even some of the small surface barriers -- see Table SI).
Because they are rare, these reconfigurations effectively divide the trajectories of each particle into distinct dipole fluctuation regions, each with different average dipole values.

\begin{figure}[h]
    \centering
    \includegraphics[width=\linewidth]{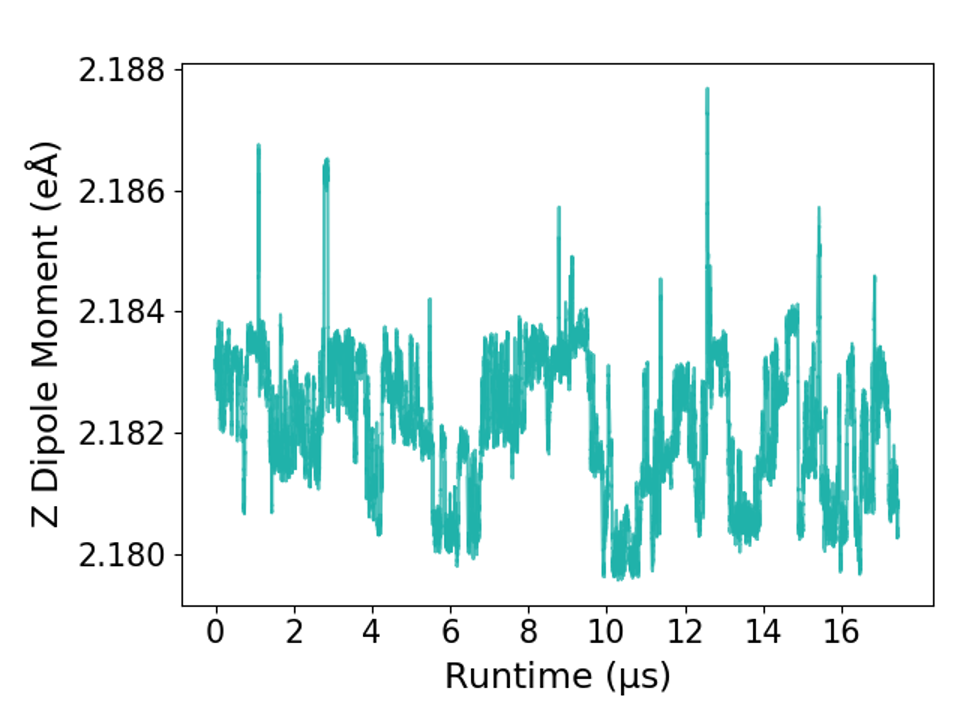}
    \caption{10,000-frame moving average of the summed system dipole from a 0.71 ML coverage trajectory run at 25 K. Multiple average dipole states emerge across the course of the simulation. This set of fluctuations gives rise to $1/\omega$ features in the corresponding $S_{\mu}(\omega)$ spectra.}
    \label{fig:multi-level-fluctuator}
\end{figure}

This collection of dipole fluctuations among the system's particles aggregate to produce multiple average dipole states in the summed system dipole time series data, an example of which can be seen in Figure \ref{fig:multi-level-fluctuator}. These multi-level fluctuators give rise to $1/\omega$ features in the corresponding dipole-dipole fluctuation spectra in Figure \ref{fig:submono_spectra}.

At higher temperatures, the \ch{CH4}-\ch{CH4} interactions become smaller relative to each particle's kinetic energy. This leads to the clusters losing their cohesiveness (see Figure \ref{fig:monolayer_cov}) as particles more commonly move across the surface to new global positions. Instead of sampling different dipole regimes in different adsorbate cluster configurations (as in the low-temperature case), the particles now freely sample the whole surface with minimal hindrance (since the adsorbate surface barriers are not significant -- see Table SI). 
Unlike with the rare collective movements observed at low temperatures, this situation involves more common but substantially less correlated particle motions. As stated earlier, trajectories that randomly sample a single distribution produce white noise; thus, it is not surprising that these fluid surfaces give rise to featureless spectra. The magnitude of the high-temperature white noise increases with temperature because more of the higher-energy $z$ positions (and thus larger dipoles) are accessed by the particles at higher temperatures.
\begin{table}
    \centering
    \renewcommand\arraystretch{1.25}
    \renewcommand{\tabcolsep}{2.5pt}
    \begin{tabular}{c|c|c|c|c|c}
        \textbf{Temp.} & \textbf{Clusters} & \textbf{Surf. Moves} & \textbf{Corr.} & \textbf{$1/\omega$} & \textbf{$1/\omega^2$} \\
        \hline
            20-40 K & Strong & Rare & High & 10$^6$-10$^9$ Hz & 10$^6$-10$^8$ Hz \\
        $>$ 50 K & Weak & Common & Low & Not Seen & Not Seen
    \end{tabular}
    \caption{Summary of submonolayer results.}
    \label{tab:subml_results}
\end{table}
Table \ref{tab:subml_results} summarizes the trends discussed for the low-temperature and high-temperature submonolayer spectra.


Identifying correlated motions is not always straightforward, as they come in many varieties (as can be seen in Figure \ref{fig:collective_motions_panel}). We used a number of methods to identify relevant motions. Tracking the number of neighbors each particle can identify particles migrating around or approaching cluster edges. Examining how cluster axial angles vary can identify cluster rotations. However, the most reliable way of identifying meaningful movements was to parse dipole moment time-series data and look for substantial changes in the average dipole moment, such as that shown in Figure S17.
The averaging step is critical, as the dipole data is very noisy from timestep to timestep.

It is important to emphasize that the frequency of the movements discussed may vary from run to run. Because these events are rare, sampling sizes and stochasticity will invariably play a role in how often they happen in a (relatively short) simulation, and such effects will be felt more greatly at lower temperatures. These effects will in turn influence the corresponding feature frequencies and magnitudes in the dipole-dipole fluctuation spectra. Nevertheless, these results illustrate how a series of collective motions that are sufficiently rare (due to temperature, cluster cohesion, and weak-but-existent surface barriers) can produce $1/\omega$ scaling at low frequencies on a model trap surface.

\subsection{Supermonolayer Dynamics and Spectra \label{superdiscussion}}

We next examine the spectra produced by adsorbate dynamics at monolayer and supermonolayer coverages. As Figure S14
illustrates, monolayer trajectories produce featureless spectra. With the surface fully covered, there is little room for significant surface movements.  As with the submonolayer coverages, some particles reside in higher-energy atop sites due to the competition between the surface-adsorbate and OPLS interactions, but unlike the submonolayer adsorbates, the monolayer adsorbates do not move significantly and thus cannot significantly contribute to the noise or dipole-dipole fluctuation spectra.

\begin{figure*}
\centering
\includegraphics[width=\linewidth]{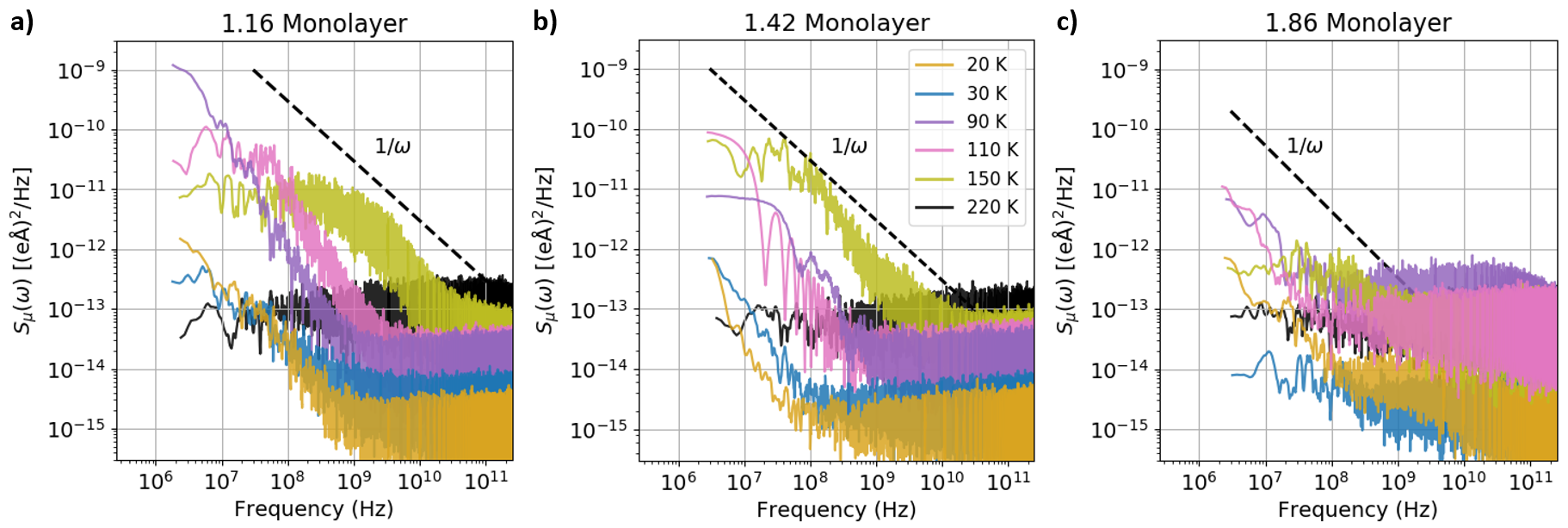}
\caption{$S_{\mu}(\omega)$ spectra obtained from three different supermonolayer coverages over a range of temperatures. The spectra were smoothed over a window of five timesteps.}
\label{fig:superML}
\end{figure*}

The spectra produced by supermonolayer dynamics, on the other hand, show several sets of features, as seen in Figure \ref{fig:superML}. At low temperatures ($<$ 60 K), spectra display scalings of $\omega^{-1}$ mixed with some $\omega^{-2}$ until around 35 K, when the spectra become nearly flat (see also Figure S15).
At higher temperatures ($>$ 70 K), however, a new set of larger-magnitude TLF-like features supersede the low-temperature features. These features increase in noise magnitude until roughly 90-110 K (depending on the coverage) before declining with a corresponding increase in cutoff frequency. These high-temperature supermonolayer features cover a greater range of noise magnitudes and frequencies than any of the other spectral features we found.

\subsubsection*{Low-Temperature Supermonolayer Dynamics}

\begin{figure*}
    \centering
    \includegraphics[width=5.5in]{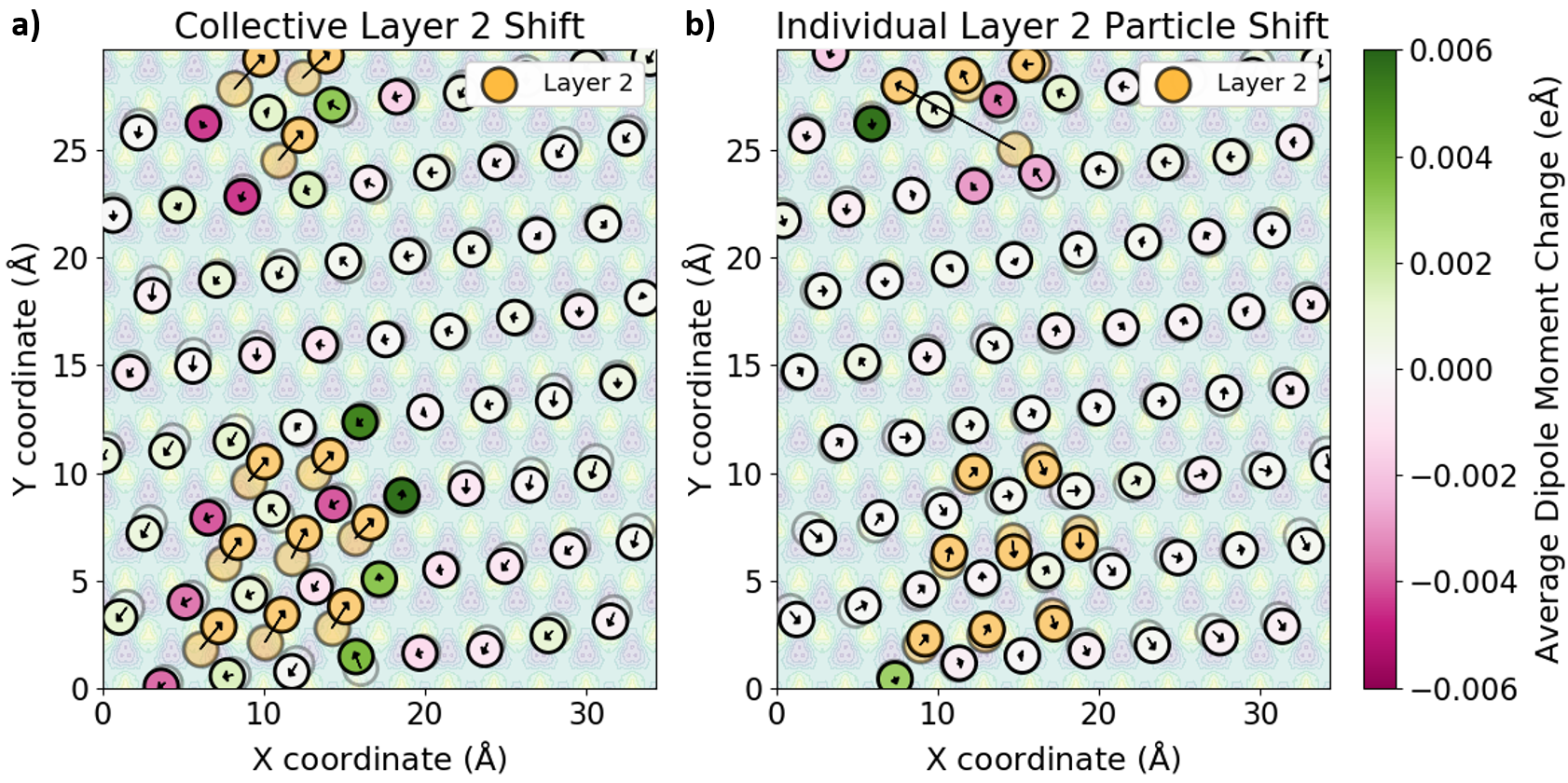}
    \caption{Depiction of \textbf{(a)} collective and \textbf{(b)} individual second layer particle movements and their effect on the average dipoles of surface adsorbates. Both examples are taken from segments of the 1.16 ML trajectory at 20 K. Arrows point from each particle's initial position (represented by the dimmed circles) to its final position. Here, only first layer particles are colored with the colormap; second layer particles, which exhibit virtually no dipole variation (see Figure \ref{fig:dip_vs_Z}), are colored yellow. As second layer particles move, they significantly impact nearby first layer particles, especially those that gain or lose second layer neighbors. In this way, the movement of particles in the unfilled second layer can cause surface dipole changes and position shifts.}
    \label{fig:low-T-superML-motions}
\end{figure*}
\begin{figure}
    \centering
    \includegraphics[width=\linewidth]{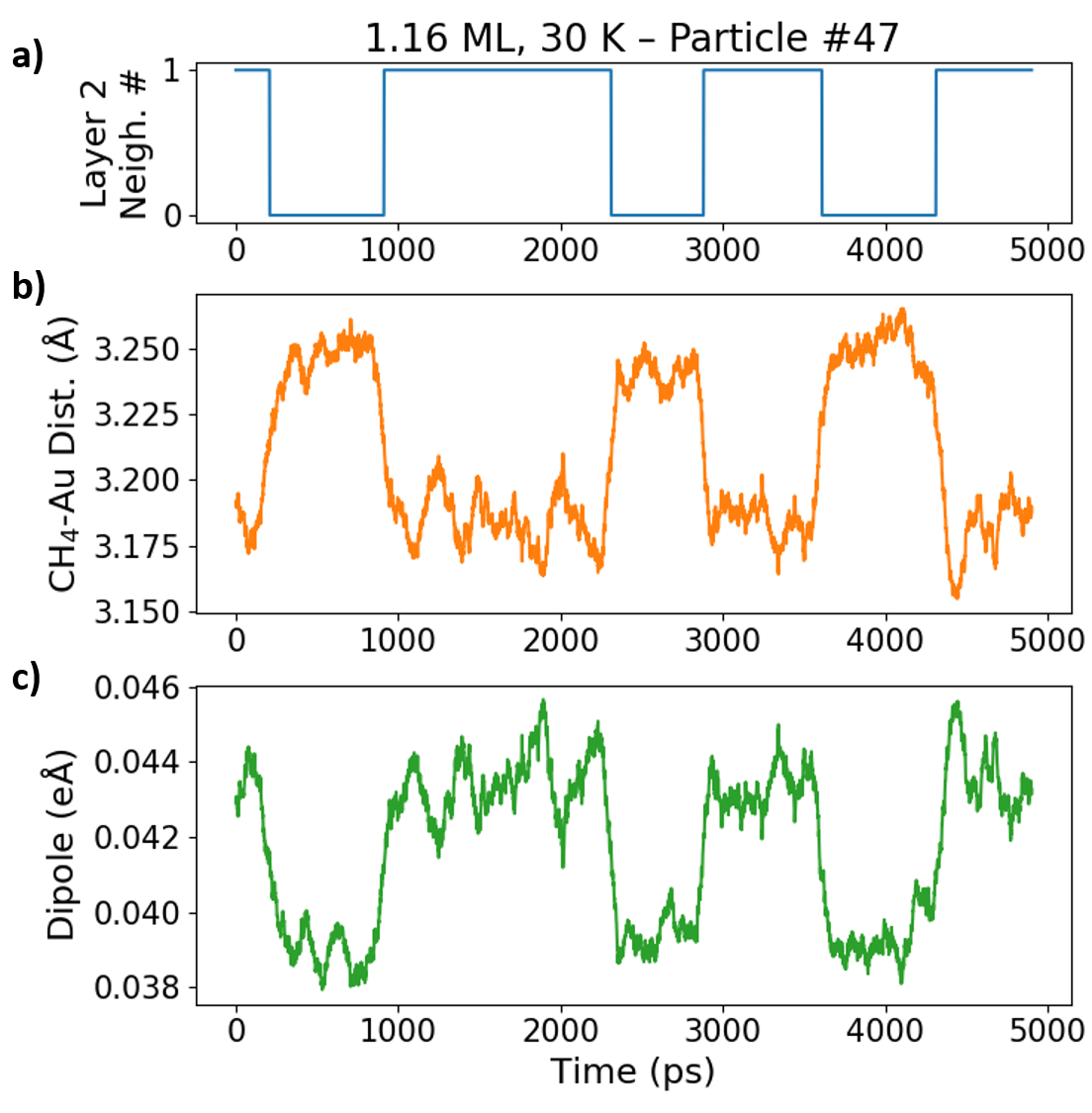}
    \caption{An example of the effect of second layer neighbors on surface adsorbates, taken from a segment of the 1.16 ML trajectory at 30 K. \textbf{(a)} Number of second layer particles neighboring Particle \#47. \textbf{(b)} 100-frame rolling average of Particle \#47's distance from the surface. \textbf{(c)} 100-frame rolling average of Particle \#47's dipole moment. When a second layer particle migrates to Particle \#47's neighborhood, it pushes the particle to a distinctly different average dipole regime.}
    \label{fig:layer2_pushing}
\end{figure}

At low temperatures, adsorbates on filled supermonolayer surfaces are mainly static (apart from in-place rattling), but they can be affected by second layer adsorbates in ways that lead to individual and collective first layer motions (see Figure \ref{fig:low-T-superML-motions}). Even a stationary second layer affects first layer particles: adsorbates with more neighboring second layer particles (at most three, as can be seen in Figure \ref{fig:coverage_panel}b) are pushed down to lower surface heights than those with no such neighbors (see Figure \ref{fig:layer2_pushing}). But, second layer particles can and do migrate around the unfilled second layer to different global positions. Similar to the submonolayer case, such movements happen more often with increasing temperature, but second layer particles are even more weakly held by the surface potential than surface particles despite experiencing OPLS interactions from first layer particles (see Figure S3)
and thus migrate more frequently at lower temperatures. As these second layer particles migrate -- at rare intervals on the ps time scale but plenty of times over the course of a $\mathrm{\mu}$s-scale trajectory -- they end up influencing different sets of first layer particles at different points in the trajectory, thus giving rise to different average surface height (and, with them, average dipole) regimes for those surface particles at those times (see Figure \ref{fig:low-T-superML-motions}). When enough particles experience these second layer induced changes, the total system dipole is moved into a different regime as well, leading to features in the dipole-dipole fluctuation spectra.

Supermonolayer surfaces can also undergo collective motions as a result of second layer particle migrations that perturb the surface in the right way, as can be seen in the examples in Figure \ref{fig:low-T-superML-motions}. However, these motions only seem to consist of slight cluster shifts and ripples rather than the diversity of submonolayer collective motions highlighted by Figure \ref{fig:collective_motions_panel}; this is likely due to submonolayer surfaces having far more room to maneuver than filled-up supermonolayer surfaces.

\begin{table}
    \centering
    \renewcommand\arraystretch{1.25}
    \renewcommand{\tabcolsep}{4pt}
    \begin{tabular}{c|c|c}
        \textbf{Motions} & \textbf{Coverages} & \textbf{Avg. Dipole Changes} \\
        \hline
        Collective Surface Motions & $<$1 ML & $<$5 me\r{A} 
        \\
        Layer 2 Pushing & $>$1 ML & 3-10 me\r{A} 
        \\
        Surface Shifts & $>$1 ML & $<$3 me\r{A} \\
        Layer Exchanges & $>$1 ML & 38-48 me\r{A} 
        \\
    \end{tabular}
    \caption{Summary of motions and the corresponding range of average per-particle dipole moment magnitude changes that they cause.}
    \label{tab:tlf_transition_magnitudes}
\end{table}

\begin{figure}
    \centering
    \includegraphics[width=\linewidth]{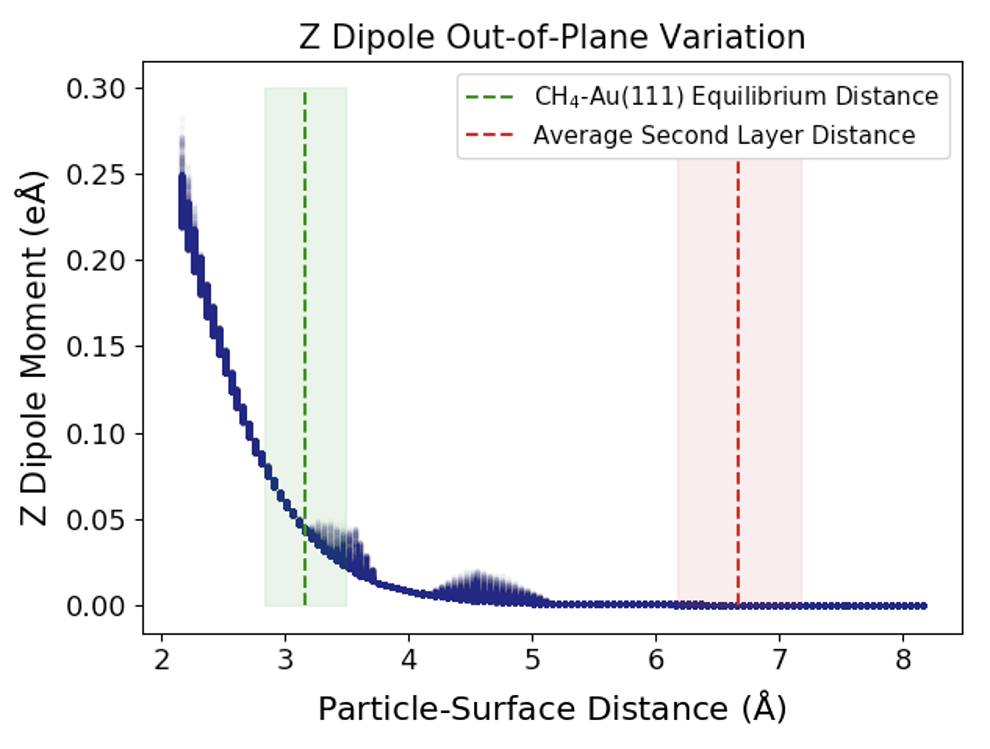}
    \caption{Plot of all possible particle $z$ dipole moments (across all $x$ and $y$ surface positions) versus the particle's distance from the electrode surface. The dark blue shading indicates how often the (dipole, distance) pair is generated from the interpolation script at different $(x,y)$ values for that distance; the bumps around 3-3.75 and 4-5 \r{A} are from dipole sensitivity to $(x,y)$ changes at those distances. The two layers are marked with mean surface distances (dotted lines) and 95\% confidence intervals (shaded regions) calculated from supermonolayer particles at 60 K. The first layer experiences a range of dipole moment changes from small shifts in any direction. Although the second layer contains virtually no dipole variation, the change in dipole moment from going between layers is an order of magnitude greater than any dipole changes within the first layer.}
    \label{fig:dip_vs_Z}
\end{figure}

Although second layer particles are responsible for first layer dipole regime changes, their own dipole changes are minimal and do not contribute to overall noise levels. Figure \ref{fig:dip_vs_Z} shows why this is: at typical second layer distances from the surface, adsorbate dipole moments are nearly zero and have virtually no variation. Although this makes movement within the second layer noiseless, it makes movement between layers significant in dipole terms, as the next section will detail.

\subsubsection*{High-Temperature Supermonolayer Dynamics}

Starting at intermediate temperatures (between 70 and 90 K), the supermonolayer spectra diverge more decisively from the submonolayer spectra. Whereas submonolayer spectra do not contain lower frequency noise features at higher temperatures, supermonolayer spectra demonstrate large low-frequency shoulders orders of magnitudes larger than any of the submonolayer low-frequency features. The emergence of such shoulders indicates that there is an additional, larger noise source present at high temperatures that is absent at lower coverages.
\begin{figure}
    \centering
    \includegraphics[width=\linewidth]{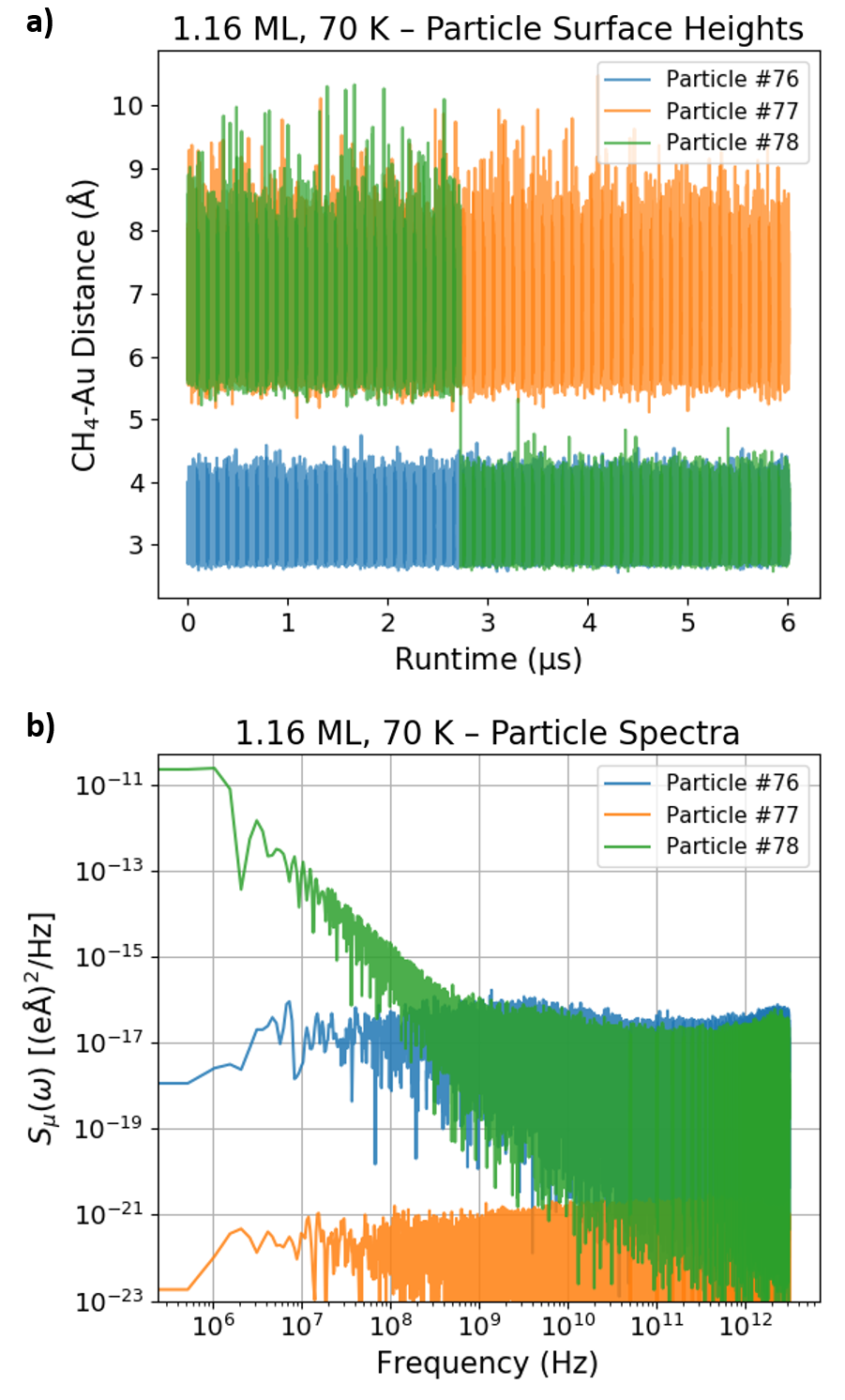}
    \caption{\textbf{(a)} Surface-adsorbate distances of selected adsorbates taken from the 1.16 ML trajectory at 70 K. Adsorbate \#78 undergoes a layer exchange that alters its surface-adsorbate distance around 2.75 $\mathrm{\mu}$s, while adsorbates \#76 and \#77 remain in their original layers. \textbf{(b)} Corresponding single-particle dipole-dipole fluctuation spectra of the selected adsorbates.}
    \label{fig:LE_indiv_particles}
\end{figure}
It turns out that this noise stems from movements of adsorbates \textit{between} layers, which we refer to as layer exchanges. Unlike the submonolayer motions in Figure \ref{fig:collective_motions_panel}, layer exchanges occur in the span of a few ps, and they have a larger impact on particle and system dipoles (see Table \ref{tab:tlf_transition_magnitudes}).

Figure \ref{fig:LE_indiv_particles} demonstrates how layer exchanges translate into the larger spectral noise observed. Of the three particles depicted, only the one that exchanged layers produces any low frequency features. Figure \ref{fig:dip_vs_Z} shows why layer exchanges produce more noise than any surface adsorbate motions. When a particle transitions from the first to the second layer, its dipole moment drops from roughly 0.044 to nearly zero e\r{A}, which is a drop several orders of magnitude larger than any fluctuation in the dipole moment due to first layer surface motions alone (and compared to the motions shown in Figure \ref{fig:collective_motions_panel}).

With large dipole changes and long mean state residence times, layer exchanges can produce significant low frequency TLF noise in the spectra of individual particles (per Equation \eqref{eqn:TLF}). However, the spectra produced by the system's total dipole moment will only be affected if layer exchanges change the makeup of the surface, where the most significant induced dipole moments are possible (see Figure \ref{fig:dip_vs_Z}). Therefore, only \textit{unidirectional} layer exchanges, rather than a set of reciprocal exchanges, will contribute to the spectra seen in Figure \ref{fig:superML}.



\begin{figure}
    \centering
    \includegraphics[width=\linewidth]{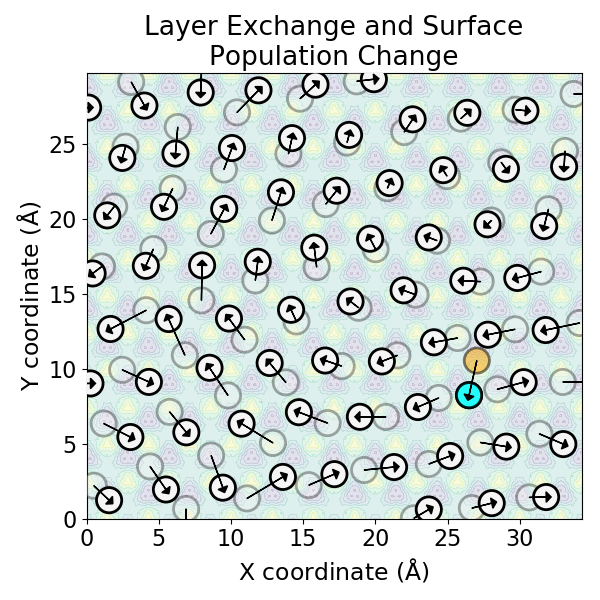}
    \caption{Layer exchange that results in a surface population change, taken from a 1.16 ML trajectory segment at 70 K. The yellow circle denotes the second layer particle before exchanging layers, while the blue circle shows the same particle after descending to the first layer.}
    \label{fig:layer-exch-with-l1-pop-change}
\end{figure}

As noted in Section \ref{sec:ads_surface_geometries}, our simulation cell can support 68- and 70-adsorbate surfaces in addition to the previously defined monolayer surface of 69 adsorbates. Second-layer OPLS interactions can make less-favorable monolayer surface populations more accessible as well. Unidirectional layer exchanges act to transition the system between these different surface populations. An example of this is shown in Figure \ref{fig:layer-exch-with-l1-pop-change}. It is this effect of losing or gaining a particle (and its large induced dipole) from the surface that leads to the system state fluctuation that produces features in the spectra in Figure \ref{fig:superML}.

As with the submonolayer surface movements, the frequency of layer exchanges may vary from run to run. With mean residence times in the microseconds for some runs, the spectra can be sensitive to additional exchange events or a change in the time that they occur. For this reason, these results are not meant to precisely recreate experimentally-observed spectra. Nevertheless, these results illustrate how thermally-activated layer exchanges can give rise to $1/\omega$ scaling regions at low frequencies on a model trap surface.


\subsubsection*{Layer Exchanges as Two-State Thermally-Activated Fluctuations} 
\label{twolevel}

One can consider layer exchanges as a two-state activated process of the kind discussed in Section \ref{sec:TLF_overview}. Adsorbates can either inhabit Layer 1 (nonzero dipole moment) or Layer 2 (near-zero dipole moment) and can stochastically change layers abruptly. The layer exchanges in our simulations do not perfectly adhere to this paradigm, as Layer 1 particles exhibit a range of values with a variance of around $4.5\times10^{-3}$ e\r{A} rather than one single value. However, the Layer 2 particles can be represented by a state with a dipole value of zero with little loss of precision (see Figure \ref{fig:dip_vs_Z}). The typical Layer 1$\rightarrow$2 dipole difference of $4.4\times10^{-2}$ e\r{A} is an order of magnitude larger than the Layer 1 dipole variance, ensuring that the magnitudes of Layer 1$\leftrightarrow$2 transitions are outside the distribution of Layer 1 dipole values. Although layer exchanges are individual motions, this treatment can be extended to the total system dipole moment, where states can be defined by layer populations each differing by roughly $4.4\times10^{-2}$ e\r{A}.

To see how well this RTS model describes our simulated spectra, we calculated analytical spectra from Equation \eqref{eqn:TLF} based on parameters from the corresponding simulations. $\Delta I$ was set to $4.4\times10^{-2}$ e\r{A} to represent the magnitude of the typical dipole fluctuation from one adsorbate vacating or joining the surface. To estimate $\tau_0$ and $\tau_1$, we calculated the average residence times of each surface population that the system experienced. For simplicity, here we present analysis performed on transitions between two layer populations. 
In our trajectories, typical $\tau_0$ and $\tau_1$ values ranged from the 100's and 1000's of ns at 90-110 K to the 10's of ns at 150 K.

\begin{figure}
    \centering
    \includegraphics[width=\linewidth]{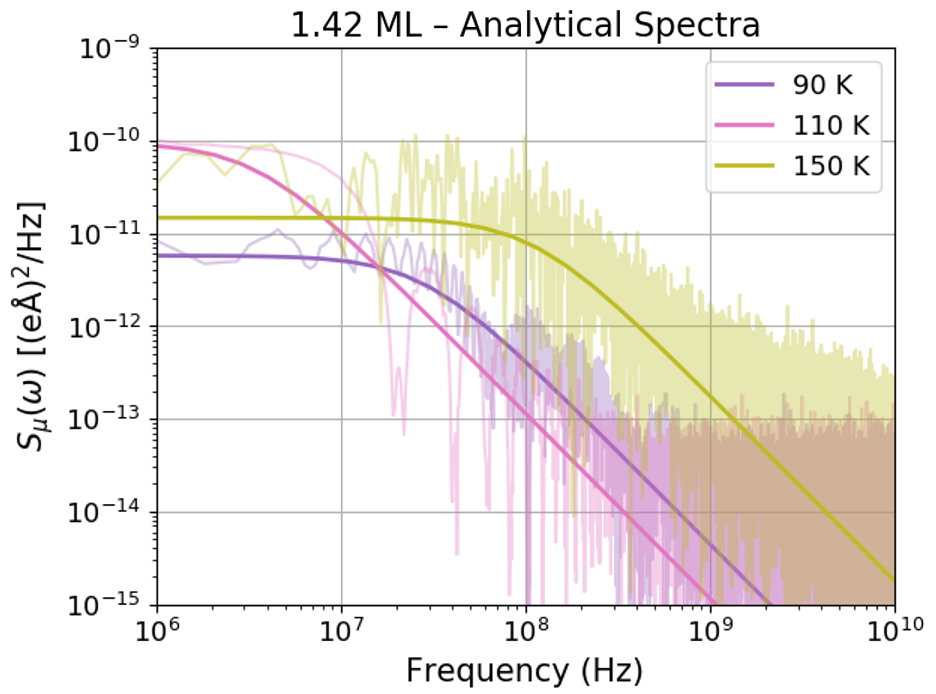}
    \caption{Comparison of simulation spectra (lighter shade) and analytical spectra calculated from the two-level fluctuator model (darker shade) for selected 1.42 ML trajectories.}
    \label{fig:superml_analytical_vs_sim}
\end{figure}

Figure \ref{fig:superml_analytical_vs_sim} compares TLF analytical spectra calculated from Equation \eqref{eqn:TLF} with spectra obtained from selected 1.42 ML trajectories. In addition, Figure S18
repeats this comparison using per-particle spectra. 
Although the simulated spectra are noisier, they follow the same frequency and magnitude trends as the analytical spectra. In particular, the relative separation between the different trajectories and the ordering of the low-frequency leveling-off is captured by the analytical spectra.


Figure \ref{fig:superml_analytical_vs_sim} and Equation \eqref{eqn:TLF} provide a rationale for why the low-frequency magnitudes of the supermonolayer spectra increase with decreasing temperature. Increasing system temperature results in more frequent layer exchanges; in RTS terms, this can be seen as decreasing \textit{both} $\tau_0$ and $\tau_1$, since the more often layer exchanges occur, the less time a particle spends in a given layer. In the low-frequency limit of Equation \eqref{eqn:TLF}, the noise magnitude is inversely proportional to the $((1/\tau_0) + (1/\tau_1))^2$ term, which means that shorter residence times will decrease the magnitude of the noise. This behavior is similar to that displayed in Figure S16.



The addition of the high-frequency white noise and two-state fluctuator regions of the spectra gives us the spectral form we see in our supermonolayer coverages above 70 K.

\section{Conclusions}
\label{Conclusions}
In this work, we have analyzed the dynamics of methane adsorbates on gold electrodes based upon first principles potentials with the aim of identifying the adsorbate motions that give rise to anomalous heating in ion traps. Based upon an extensive set of molecular dynamics simulations run at a wide range of temperatures and surface coverages, we were able to correlate adsorbate motions such as cluster reconfigurations and layer exchanges with the different frequency-dependent features of the adsorbate dipole-dipole fluctuation spectra that can ultimately be related to trap noise. Counter to prevailing intuition, we find that accounting for adsorbate-adsorbate correlations is pivotal for observing $1/\omega$ frequency scalings and that rare, fundamentally collective adsorbate motions are responsible for the $1/\omega$ features in these spectra at MHz frequencies. In contrast, single adsorbate well-rattling motions or surface diffusion led to white noise at frequencies orders of magnitude larger than those realized in ion traps. In particular, we find that the collective rotations or translations of strongly-interacting adsorbates within clusters give rise to $1/\omega$ scalings at submonolayer coverages, while layer exchanges of adsorbates between the first two adsorbate monolayers are the greatest source of these scalings at supermonolayer coverages. Key features of our simulated spectra can be reproduced by two-level fluctuator models parameterized with the intensities and residence times associated with these motions, strongly corroborating our findings. This work thus establishes rare, classical adsorbate motions as atomistic sources of anomalous heating using some of the most realistic molecular dynamics simulations undertaken to date and dispels the notion that adsorbate motions are too high in frequency to be key contributors.  

That said, our model involved a number of notable simplifications. First, our methane adsorbates were treated as classical point masses, which enabled us to treat the methane-methane interactions via an OPLS potential rather than in a fully \textit{ab initio} manner. While this is a reasonable simplification for a spherically-symmetric molecule like methane, it would be a far more severe approximation for larger hydrocarbons containing many more vibrational and rotational degrees of freedom. Second, we did not take any nuclear quantum effects into account. These could become important at the lowest temperatures explored here and would open up the possibility of the methane adsorbates more readily traversing the gold surface. While tunneling would not impact the $1/\omega$ scaling that emerges due to layer exchanges at $>$70 K, it could affect the forms of the submonolayer spectra, an issue which would have to be investigated during future studies. Moreover, our single-crystalline gold electrodes were unrealistically smooth. In practice, most electrodes will be formed from either polycrystalline materials or possess a wide range of defects. It is likely that adsorbates traversing such uneven surfaces would also generate dipole fluctuations and, consequently, trap noise. Lastly, throughout this work, we have relied upon the patch potential model to link dipole fluctuations to trap noise even though it is but one approximate model of many potential models for this noise.\cite{Kumph_2016} For these reasons, our model is not meant to precisely reproduce experimental spectra, but rather to provide insights regarding the microscopic origins of anomalous heating that we believe will generalize to even more realistic models and systems. 

Despite these simplifications, our model still reveals a variety of regions with $1/\omega$ scaling. The fact that it did so without involving multiple types of adsorbates or accounting for vibrational degrees of freedom suggests that the mere presence of even simple molecules on electrode surfaces can lead to anomalous heating. This points to a universal molecular origin for this phenomenon. Even so, more massive hydrocarbons involved in the trap fabrication process and that are more likely to be circulating in the ambient air, such as isopropanol, are much more likely than methane to be adsorbed onto electrode surfaces. These more massive hydrocarbons would be expected to interact more strongly with the surface and to exhibit larger, more varied dipole moments, most likely shifting the $1/\omega$ scaling regimes we observed to even lower frequencies. More strongly interacting hydrocarbons may also remedy a deficiency of our results -- the fact that we only observe $1/\omega$ scalings at temperatures well below room temperature -- by shifting the spectra to the higher temperatures employed in most experiments. Molecules with more vibrational degrees of freedom are additionally more likely to give rise to motions that can be activated in different frequency regimes, which would result in the emergence of multiple, overlapping two-level or multi-level fluctuators that can combine to yield $1/\omega$ scaling over a much larger frequency regime.\cite{Zanolla} We thus look forward to future simulations of the noise produced by a variety of interacting adsorbates with many degrees of freedom, which may hold the key to reproducing experimental spectra with even greater fidelity.\\

\section*{Data Availability}
The data that support this study, including the DFT adsorption energies and dipole moments used to compute dipole-dipole autocorrelation functions and fluctuation spectra, are available at \url{https://www.github.com/blfoulon/ion-trap-anomalous-heating-simulation/}.

\acknowledgements
The authors thank Dustin Hite, Jeremy Sage, Jonathon Sedlacek, John Chiaverini, and Hartmut H{\"a}ffner for helpful discussions. This work was performed under the auspices of the U.S. Department of Energy by Lawrence Livermore National Laboratory under Contract DE-AC52-07NA27344. We also thank the Brown Presidential Fellowship Program for supporting Yuan Liu. Part of this research was conducted using computational resources and services at the Center for Computation and Visualization, Brown University. 
\\

B.R. and V.L. oversaw this work equally.  

\normalem

\bibliography{ref}{}
\end{document}


\title{Supplemental Information for ``How Correlated Adsorbate Dynamics on Realistic Substrates Can Give Rise to 1/$\omega$ Electric-Field Noise in Surface Ion Traps"}

\author{Benjamin Foulon}
\affiliation{Department of Chemical Engineering, Brown University, Providence, RI 02912, USA}
\author{Keith G. Ray}
\affiliation{Quantum Simulations Group, Lawrence Livermore National Laboratory, Livermore, CA 94550, USA}
\author{Changeun Kim}
\affiliation{Quantum Simulations Group, Lawrence Livermore National Laboratory, Livermore, CA 94550, USA}
\author{Yuan Liu}
\affiliation{Center for Ultracold Atoms, Research Laboratory of Electronics, Massachusetts Institute of Technology, Cambridge, MA 02139, USA}
\author{Brenda M. Rubenstein}
\affiliation{Department of Chemistry, Brown University, Providence, RI 02912, USA}
\author{Vincenzo Lordi}  
\affiliation{Quantum Simulations Group, Lawrence Livermore National Laboratory, Livermore, CA 94550, USA}

\maketitle

\section{Checks on Parameter Convergence in Molecular Dynamics Simulations}

\begin{figure}[h]
\centering
\includegraphics[scale=0.45]{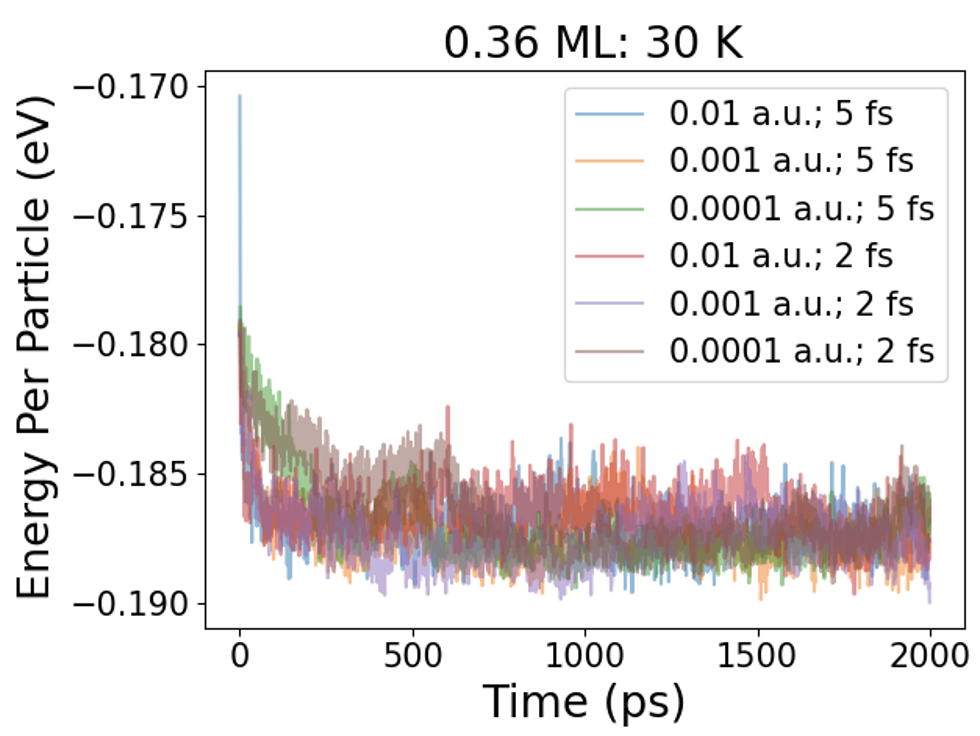}
\caption{Effects of varying the Langevin coefficient (reported in a.u., where 1 a.u. $= 4.13\times 10^{16}$ s$^{-1}$) and the MD time step (reported in fs) on energy conservation and fluctuations during the course of an MD simulation for 0.36 monolayer coverage at 30 K. A Langevin coefficient of 0.01 au and an MD time step of 5 fs were ultimately employed in our simulations.} 
\label{supp_fig:subML-params}
\end{figure}

\begin{figure}[h]
\centering
\includegraphics[scale=0.45]{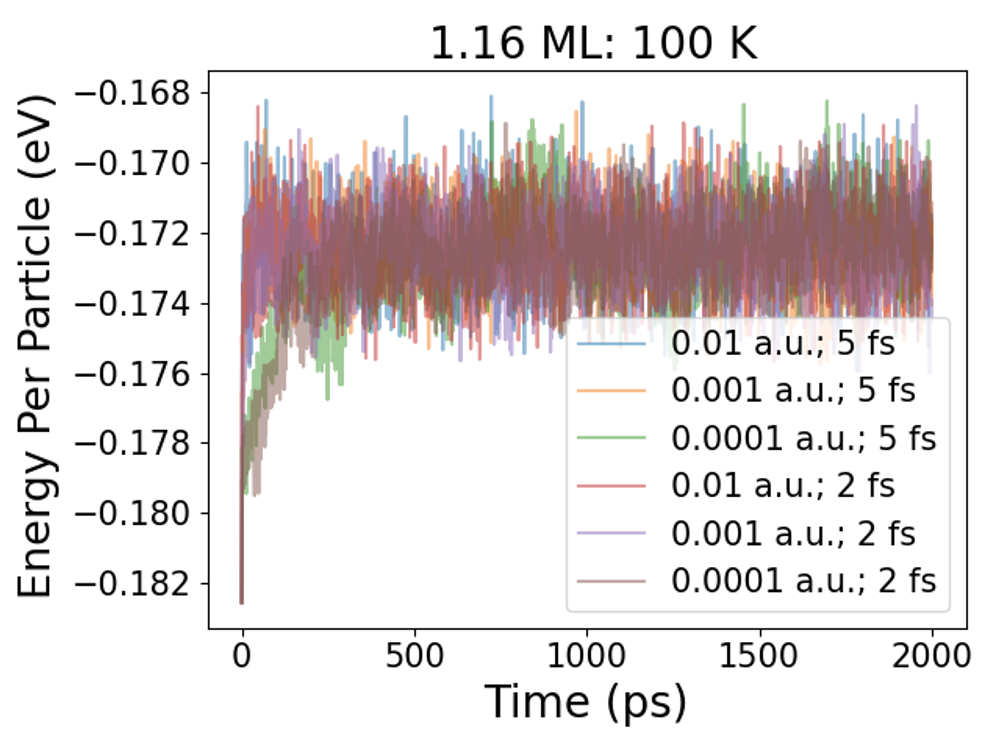}
\caption{Effects of varying the Langevin coefficient (reported in a.u., where 1 a.u. $= 4.13\times 10^{16}$ s$^{-1}$) and the MD time step (reported in fs) on energy conservation and fluctuations during the course of an MD simulation for a 1.16 monolayer coverage at 100 K. A Langevin coefficient of 0.01 au and an MD time step of 5 fs were ultimately employed in our simulations.}
\label{supp_fig:superML-params}
\end{figure}

\section{Methane-Gold Potential Energy Surface}

\begin{table}[h]
    \centering
    \renewcommand\arraystretch{1.25}
    \renewcommand{\tabcolsep}{8pt}
    \begin{tabular}{c|c|c}
        \textbf{Barrier Type} & \textbf{Barrier Height (K)} & \textbf{Barrier Height (eV)} \\
        \hline
           Surface Atop-Bridged & 0.35 & $3 \times 10^{-5}$ 
        \\
        Surface Atop-Hollow & 1.16 & $1 \times 10^{-4}$ 
        \\
        Surface Bridged-Hollow & 1.97 & $1.7 \times 10^{-4}$ 
        \\
        Surface Hollow-Bridged & 24.37 & $2.1 \times 10^{-3}$ 
        \\
        Surface Bridged-Atop & 35.16 & $3 \times 10^{-3}$ 
        \\
        Surface Hollow-Atop & 58.37 & $5 \times 10^{-3}$
        \\
    \end{tabular}
    \caption{Surface Potential Barrier Heights}
    \label{supp_tab:barrier_heights}
\end{table}

\begin{figure}[h]
    \centering
    \includegraphics[width=4.0in]{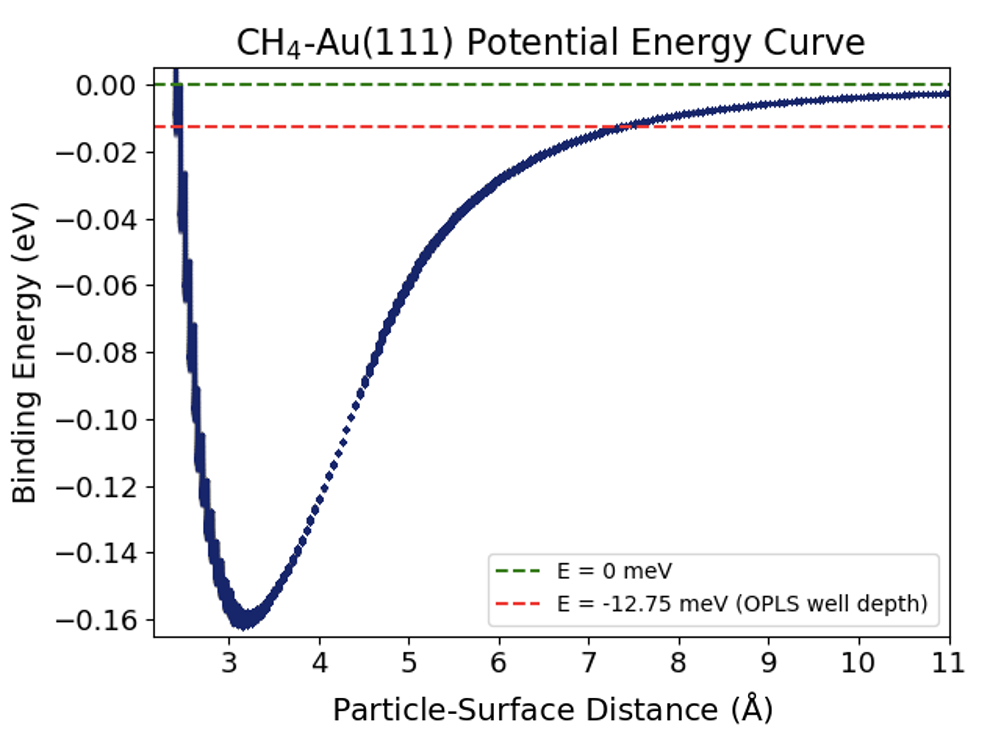}
    \caption{Adsorbate-surface potential (binding energy)  versus methane distance from the gold electrode surface. A range of values are reported for each particle-surface distance because adsorbates can reside at a variety of $(x,y)$ surface positions, each of which have their own binding energies. Such site-specific variations are shown in Figure 1 
    at the equilibrium binding distance, where these variations are greatest.}
    \label{supp_fig:PE_vs_Z}
\end{figure}
\section{Additional Simulation Parameters}
\begin{table}[h]
    \centering
    \renewcommand\arraystretch{1.25}
    \renewcommand{\tabcolsep}{4pt}
    \begin{tabular}{c|c}
        {\bf Coverage ($N_{\ch{CH4}}$)} & {\bf Runtime ($\mathrm{\mu}$s)}\\
        \hline
        0.36 ML (25) & 19 \\
        0.52 ML (36) & 17\\
        0.71 ML (49) & 17\\
        1.0 ML (69) & 4\\
        1.16 ML (80) & 9-12\\
        1.42 ML (98) & 6-7\\
        1.86 ML (128) & 6-9\\
    \end{tabular}
    \caption{Runtimes of the spectra featured in this work.}
    \label{supp_tab:runtimes}
\end{table}
\section{Time Correlation Functions That Underlie Dipole-Dipole Fluctuation Spectra}

One challenge associated with computing trap noise spectra based upon dipole-dipole correlation functions is deciding what form the correlation functions should assume. Given that our dipole data is two-dimensional ($\mathcal{T} \times N$), we have two broad options. If we decided we wanted to compute the autocorrelation function of the \textit{summed} total dipole moment of all adsorbates, we would use the following expression:

\begin{equation}
\label{supp_eqn:cfxn_biased}
    C_{\mu,\mu}(\tau) = \frac{1}{(\mathcal{T}-\tau)} \sum_{k=1}^{\mathcal{T}-\tau} \left[ \bigg(\sum_{i}^{N} \big(\mu_{i,z}(t_k)\big) - \langle \mu_z \rangle\bigg)\bigg(\sum_{j}^{N} \big(\mu_{j,z}(\tau+t_{k})\big) - \langle \mu_z \rangle\bigg) \right].
\end{equation}
If we instead wanted to first compute the dipole autocorrelation functions of each \textit{individual} particle and then sum them up, we would use the following expression:

\begin{equation}
\label{supp_eqn:cfxn_biased2}
    C_{\mu,\mu}(\tau) = \frac{1}{(\mathcal{T}-\tau)} \sum_{k=1}^{\mathcal{T}-\tau} \left[ \sum_{i}^{N} (\mu_{i,z}(t_k) - \langle \mu_{z} \rangle) (\mu_{i,z}(\tau+t_{k}) - \langle \mu_{z} \rangle) \right].
\end{equation}
In the above equations, $N$ is the number of particles and $\mathcal{T}$ is the number of time steps used in the simulation. As these equations suggest, another thing to consider is what dipole averages are appropriate to use in these expressions. For the summed case, using the time-averaged summed system dipole is a straightforward choice, but for the individual case there are several options. We could choose $\langle \mu_z \rangle$ to be the ensemble average of all adsorbate dipoles in the system. Alternatively, we could choose $\langle \mu_z \rangle$ to be an $N$-length vector composed of $N$ per-particle average dipoles.

\begin{figure*}[h]
    \centering
    \includegraphics[width=0.95\linewidth]{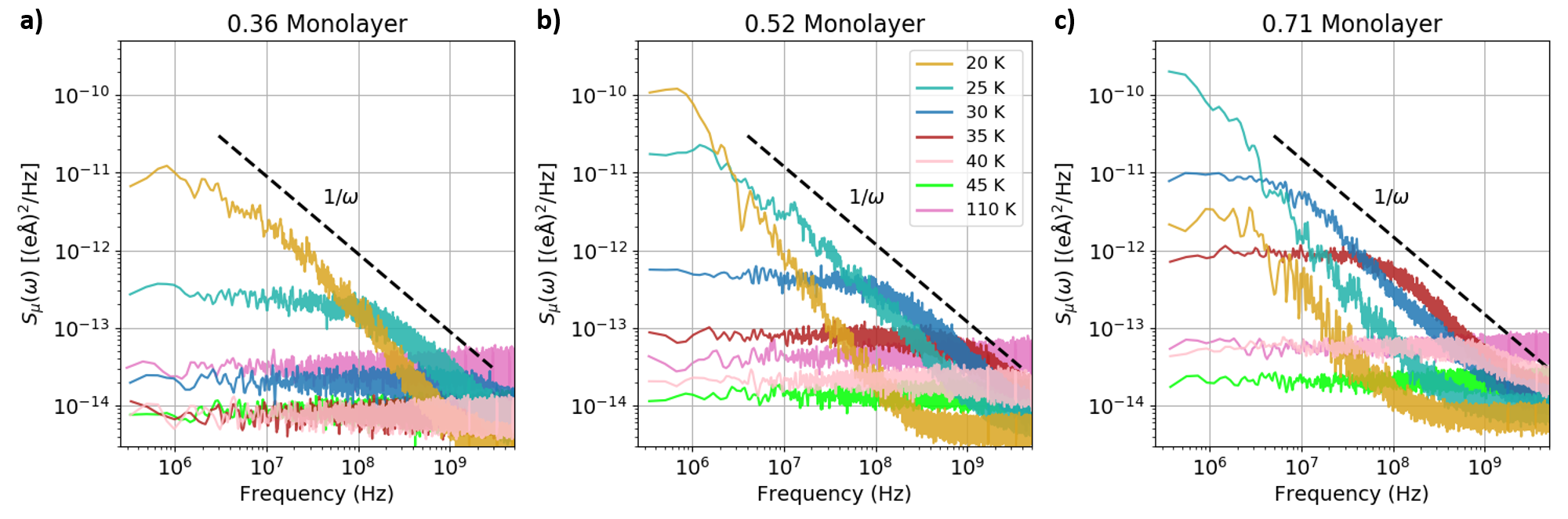}\caption{ \textit{Individual} $S_{\mu}(\omega)$ spectra obtained from dipole time series data of the submonolayer coverage runs. (Analogous to Figure 4). 
    }
    \label{supp_fig:submono_spectra_indiv}
\end{figure*}
\begin{figure*}[h]
\centering
\includegraphics[width=0.95\linewidth]{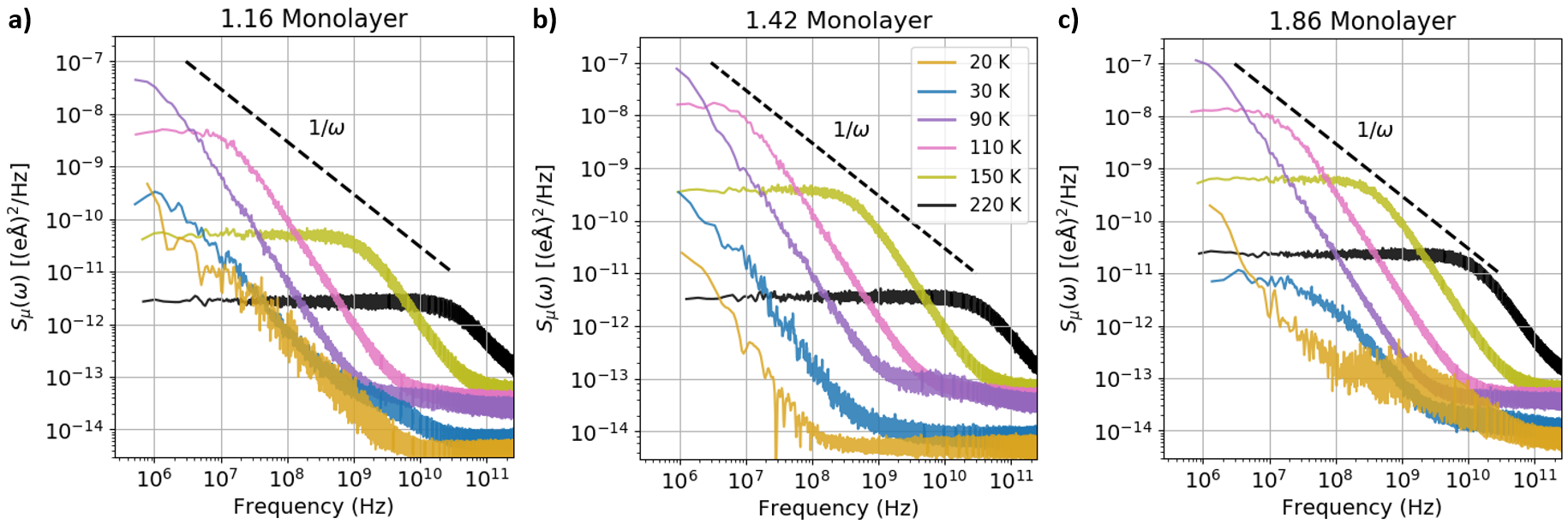}
\caption{\textit{Individual} $S_{\mu}(\omega)$ spectra obtained from three supermonolayer coverages over a range of temperatures.
(Analogous to Figure 8).
}
\label{supp_fig:superML_indiv}
\end{figure*}

To compare the different methods, Figures \ref{supp_fig:submono_spectra_indiv} and \ref{supp_fig:superML_indiv} depict dipole-dipole fluctuation spectra of the same trajectories shown in Figures 4 and 8, 
respectively, with the individual-style autocorrelation functions used instead of the summed autocorrelation functions (the choice of mean ended up not materially affecting the individual-style spectra very much). Using summed dipole correlation functions results in noisier FT spectra across all trajectories. Differences between the two methods are more apparent in the supermonolayer spectra than the submonolayer spectra. This is primarily because individual-style correlation functions capture \textit{all} layer exchanges -- since layer exchanges are individual particle motions -- and thus increases how much noise is observed in the supermonolayer spectra where they occur. On the other hand, summed correlation functions only capture layer exchanges that result in surface population changes or realignments, which effectively excludes reciprocal layer exchanges. 


We decided to use summed autocorrelation functions for the results in the body of the paper. Although noisier, system dipoles are easier to measure experimentally and do not rely on the potentially problematic labelling of indistinguishable particles. 

Given that we fixed our system center of mass to reduce drift, it is prudent to be alert for possible small complementary system-wide shifts in response to unidirectional layer exchanges. To head this off, any detected drifts can be corrected for in correlation function calculations. In practice, even if left uncorrected, such shifts would have no effect on summed dipole regime residence times (since they occur at existing events rather than constituting new events) and thus do not affect the frequency scaling or the frequencies at which different features are found in the spectra. Overall magnitudes 
would only be affected by a small scaling factor (significantly less than one logarithmic unit); this would barely budge the positioning of the spectra and is orders of magnitude less than the differences between the spectral trajectories at different temperatures.



\section{XY Dipole-Dipole Fluctuation Spectra}

\begin{figure}[h]
\centering
\includegraphics[width=7.0in]{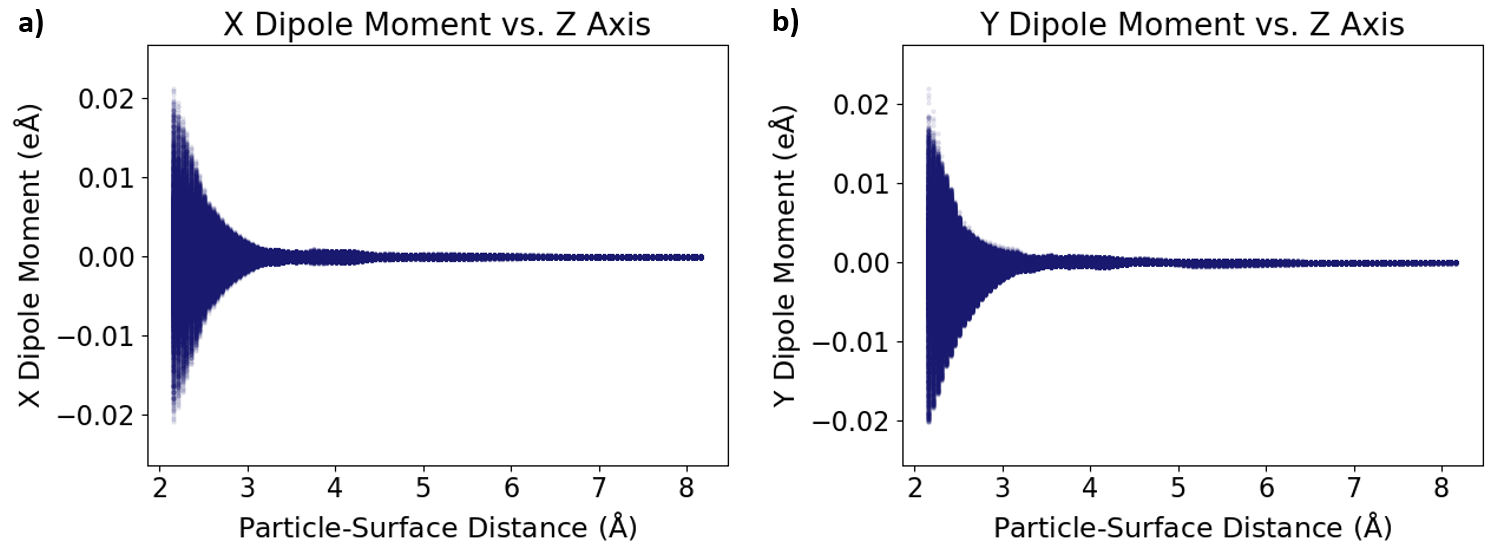}
\caption{Plot of all possible particle \textbf{(a)} $x$ and \textbf{(b)} $y$ dipole moment components (across all $x$ and $y$ values) versus the particle's distance from the electrode surface. The dark blue shading indicates how often the (dipole, distance) pair is generated from the interpolation script at different $(x,y)$ values for that distance. Comparing to Figure 11, 
it can be seen that $x$ and $y$ dipole components are consistently orders of magnitude smaller than $z$ dipole components at distances near the surfaces (while all components have very small values at second layer distances).}
\label{supp_fig:X_and_Y_dipoles}
\end{figure}

Figure \ref{supp_fig:X_and_Y_dipoles} shows the values that the $x$ and $y$ dipole components can take; they are orders of magnitude smaller than the values that the $z$ dipole components can take (see Figure 11).

\begin{figure}[h]
\centering
\includegraphics[width=\linewidth]{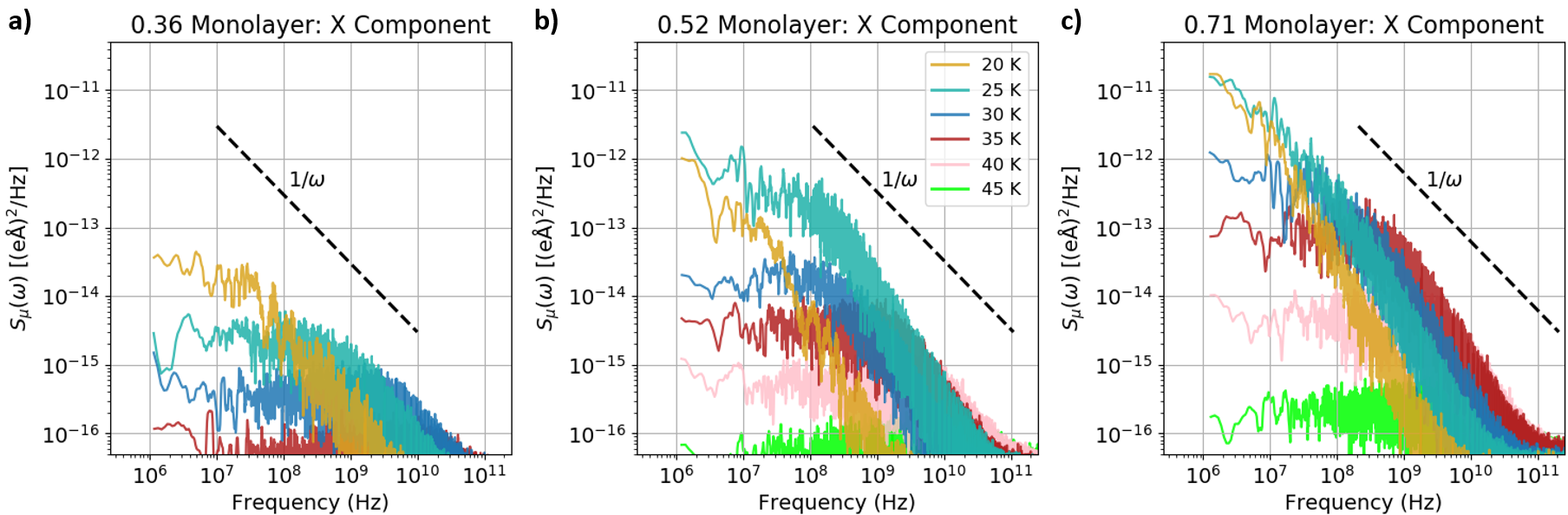}
\caption{Submonolayer spectra obtained from $x$ components of the \ch{CH4} dipoles. The spectra were smoothed over a window of five timesteps.
}
\label{supp_fig:subML-X}
\end{figure}

\begin{figure}[h]
\centering
\includegraphics[width=\linewidth]{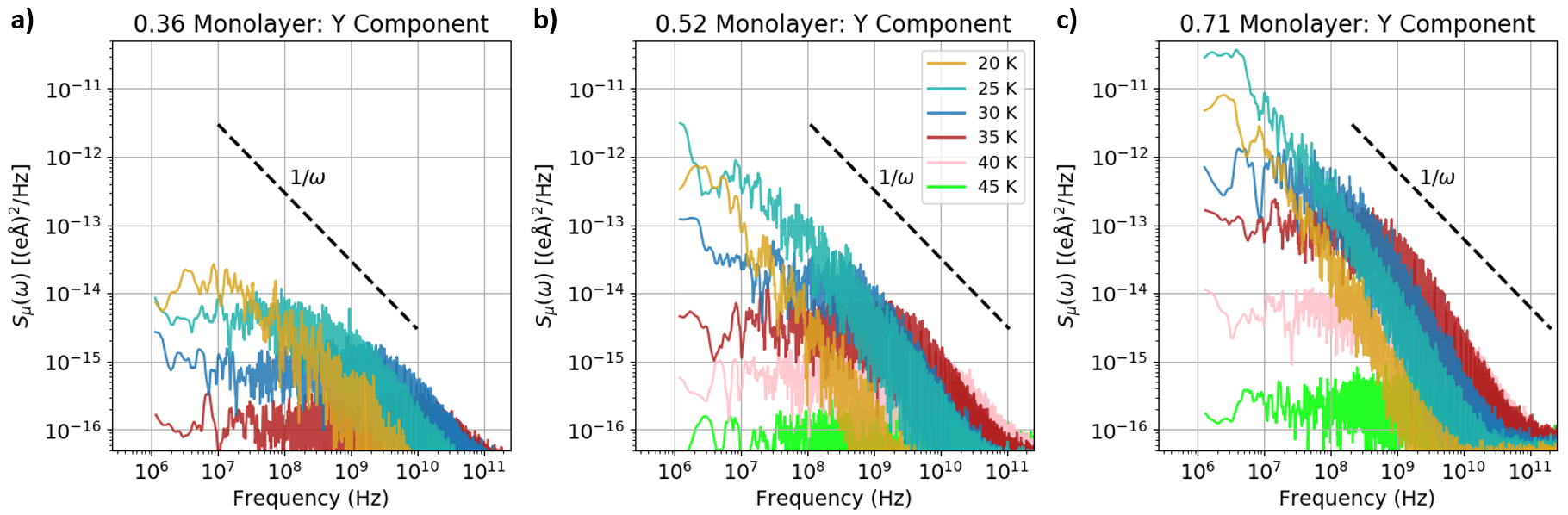}
\caption{Submonolayer spectra obtained from $y$ components of the \ch{CH4} dipoles. The spectra were smoothed over a window of five timesteps.
}
\label{supp_fig:subML-Y}
\end{figure}

Figures \ref{supp_fig:subML-X} and \ref{supp_fig:subML-Y} show fluctuation spectra using $x$ and $y$ dipole components for the submonolayer coverages. Although for most temperatures they are quite low, for the lowest of the temperatures at the 0.52 and 0.71 ML coverages the spectra take on large magnitudes that are comparable to or at times exceed the spectra generated from the $z$ dipole values. The $x$ and $y$ dipole-dipole fluctuation spectra show similar features and scalings to the $z$ dipole-dipole fluctuation spectra.

Supermonolayer $x$ and $y$ dipole-dipole fluctuation spectra are not shown here: their magnitudes are consistently below those of the $z$ dipole-dipole fluctuation spectra. At the intermediate and high temperatures at which layer exchanges occur, the magnitudes are even lower and the spectra are mostly flat, likely an indication of the central role that out-of-plane rather than in-plane variations play in the noise generated by layer exchanges. 

\section{Visualizations of Equilibrium Surface Coverages}

\subsection{Monolayer Coverages}
\label{supp_sec:monolayers}

\begin{figure*}[h]
\centering
\includegraphics[width=\linewidth]{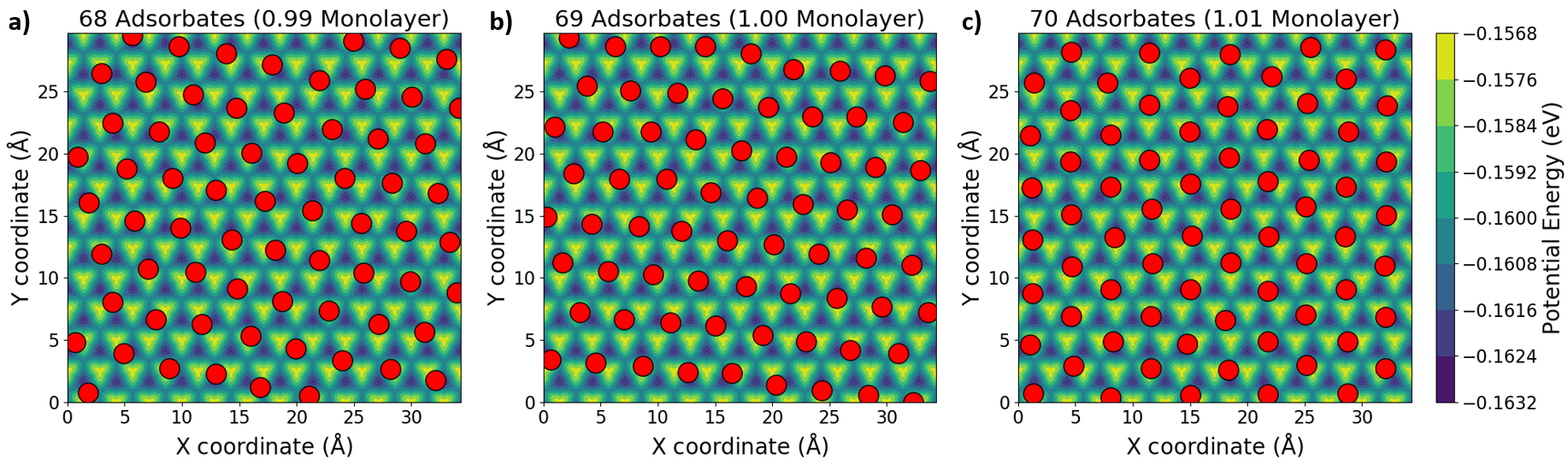}
\caption{Snapshots of adsorbate positions and configurations on the Au(111) surface at near-monolayer surface coverages: \textbf{(a)} 68 particles (0.99 ML coverage); \textbf{(b)} 69 particles (1.00 ML coverage); \textbf{(c)} 70 particles (1.01 ML coverage). Although we define $N = 69$ as our monolayer occupancy, both \textbf{(a)} and \textbf{(c)} are viable surface coverages and can manifest in supermonolayer trajectories.} 
\label{supp_fig:MLs}
\end{figure*}

\begin{figure*}[h]
\centering
\includegraphics[width=\linewidth]{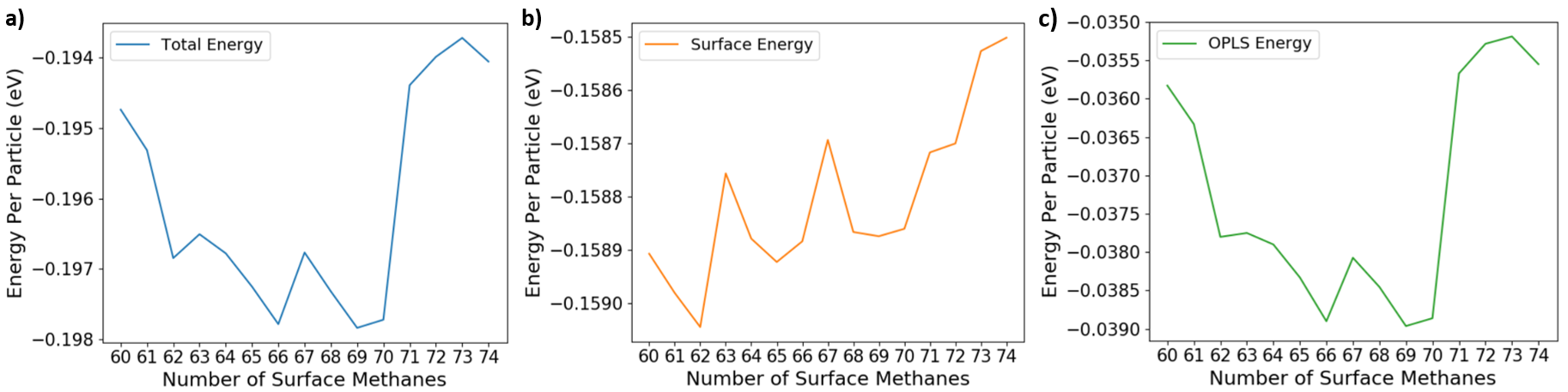}
\caption{Energy contributions to the total energy per particle as a function of the number of methanes on the electrode surface. The OPLS energy \textbf{(c)} is the main factor in determining the per-particle energy of monolayer coverages (note that the range of variation in the entire surface energy plot \textbf{(b)} is at the scale of the variation between two neighboring tick marks in the OPLS energy plot). This reflects the strength of the OPLS potential relative to $x,y$ methane-gold barriers at the surface.}
\label{supp_fig:Particle_Energy_neighbor_dist}
\end{figure*}

To arrive at our equilibrium monolayer surface coverage of 69 particles, we searched for the monolayer coverage with the lowest energy and visually inspected the related configurations to ensure they were homogeneous. Figure \ref{supp_fig:MLs} shows selected monolayer coverage visualizations, while Figure \ref{supp_fig:Particle_Energy_neighbor_dist} shows how the interparticle and surface energies vary with the number of surface adsorbates. Any attempts to add additional particles beyond 74 resulted in particles getting ejected from the surface. Although there is another local per-particle energy minimum at $N = 66$, the ability to add more surface particles (and thus lower overall system energy) pushes the size of the monolayer to 69 particles.

The OPLS energies of the cluster particles is the major factor in determining the favorability of surface coverages. Both the 68- and 70-adsorbate coverages are at low enough energies that supermonolayer coverages can transition between these states and the 69-adsorbate monolayer when the temperature is high enough for layer exchanges to occur.

\subsection{Other Coverages}

\begin{figure*}[h]
\centering
\includegraphics[width=\linewidth]{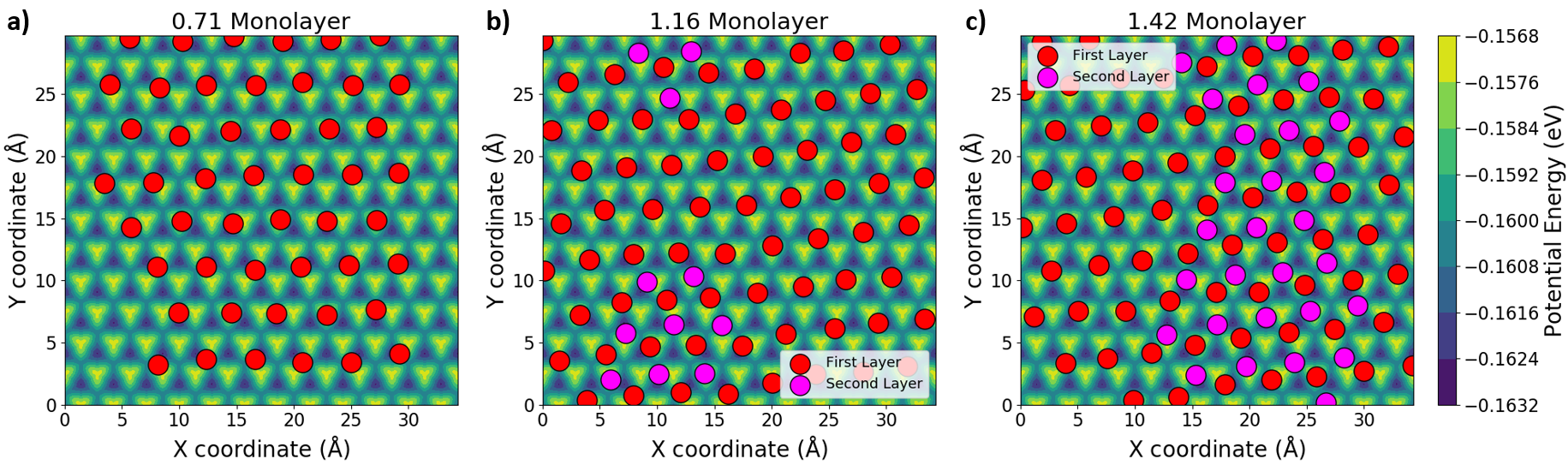}
\caption{Representative snapshots of \textbf{(a)} 49 adsorbates (0.71 ML coverage); \textbf{(b)} 80 adsorbates (1.16 ML coverage); \textbf{(c)} 98 adsorbates (1.42 ML coverage). The red and purple circles denote the particles in the first and second layers, respectively. These clearly illustrates the nearly regular pattern that emerges among the particles at sufficiently high densities, even though this pattern is incommensurate with the underlying surface potential. At sufficiently high second monolayer coverages, patterns emerge in the second layer particle configurations.}
\label{supp_fig:other_coverages}
\end{figure*}

In addition to near-monolayer coverages, we also depict two representative supermonolayer coverages (Figures \ref{supp_fig:other_coverages}b and \ref{supp_fig:other_coverages}c) to illustrate how the particles arrange themselves once the lowest-energy monolayer configuration is fully filled. In addition, Figure \ref{supp_fig:other_coverages}a depicts the highest submonolayer coverage ($N = 49$) featured in this work.

As the loosely-held second layer grows in size, it becomes more difficult to maintain its cohesion in the simulations. Higher temperatures can increase this difficulty. This can lead to second-layer particles being pushed to or drifting to surface heights beyond the second layer. We noticed that this can happen in the $N=128$ runs to a few particles as temperatures increased. This can have the effect of reducing the effective coverage from 1.86 ML to between 1.77 and 1.84 ML, depending on the number of particles. However, because the dipole variation is virtually non-existent by second layer distances, these few particles do not see their dipoles affected by this behavior.


\vfill

\pagebreak[4]

\section{Dipole-Dipole Autocorrelation Functions}

\begin{figure}[h]
\centering
\includegraphics[width=\linewidth]{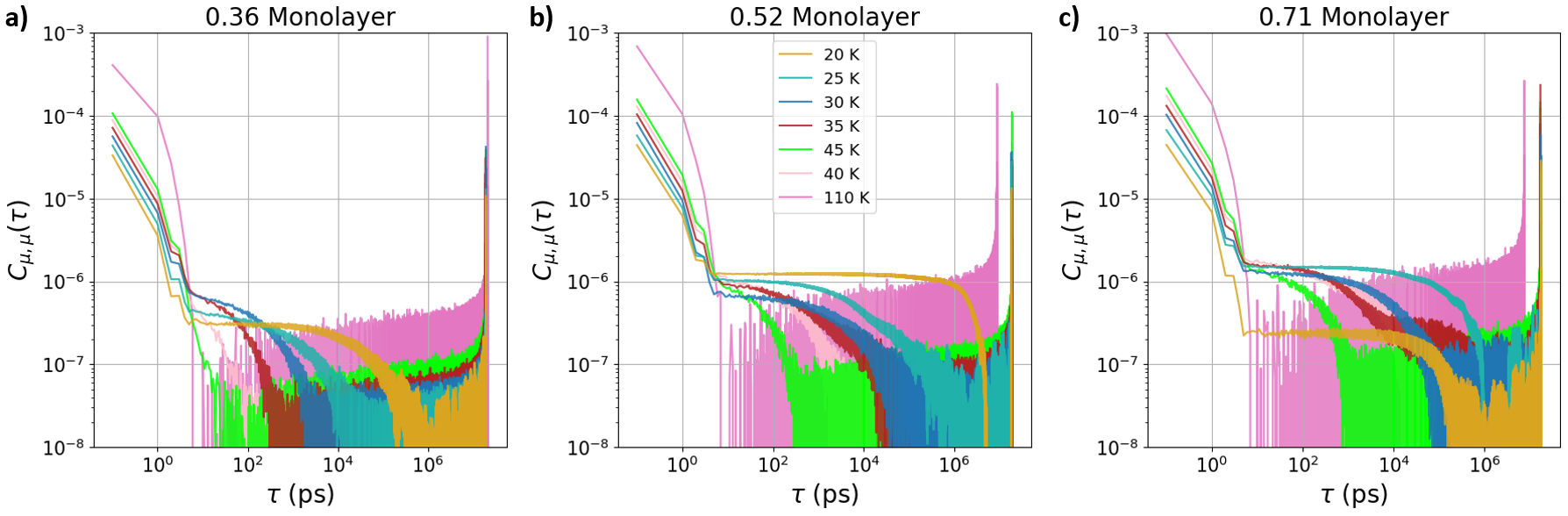}
\caption{$Z$ dipole-dipole autocorrelation functions  obtained for submonolayer coverages. The uptick at the highest of $\tau$'s is due to the quantity $(\mathcal{T} - \tau)$ approaching zero (where $\mathcal{T}$ is the number of timesteps in the trajectory). These autocorrelation functions were subsequently Fourier transformed to produce the spectra presented in the text.}
\label{supp_fig:subML_cfxns}
\end{figure}

\begin{figure*}[h]
\centering
\includegraphics[width=\linewidth]{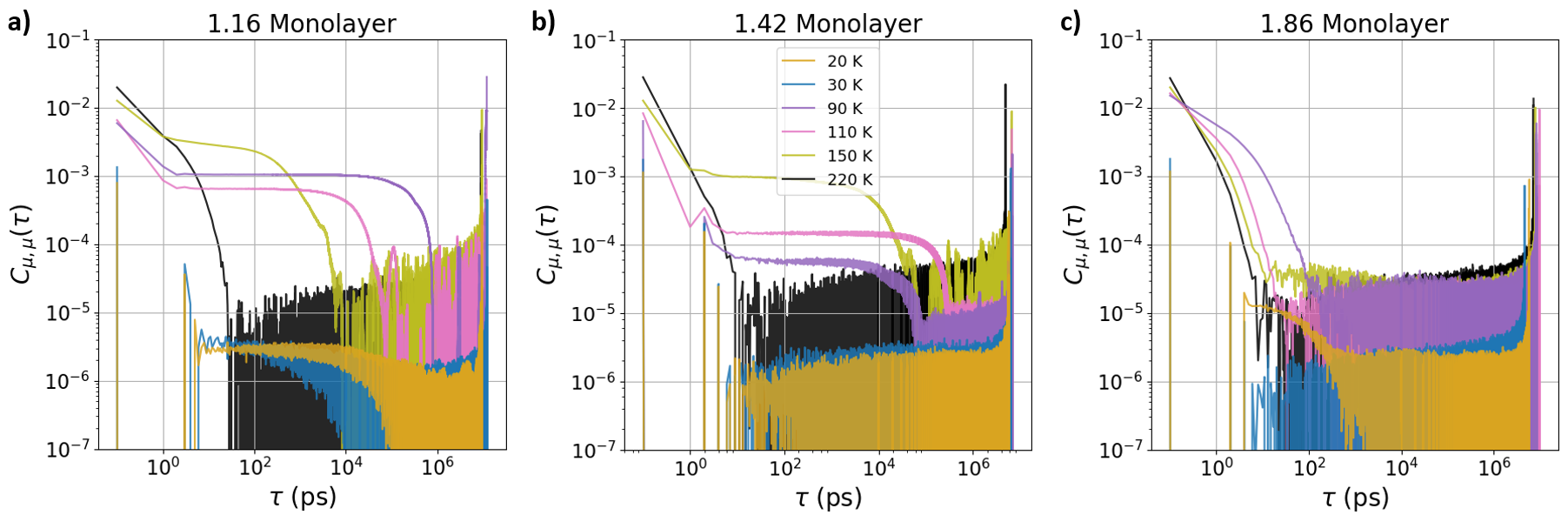}
\caption{$Z$ dipole-dipole autocorrelation functions obtained for supermonolayer coverages. Some of these functions attained negative values (off of the scale), meaning that the particles were anticorrelated, at short times. The uptick at the highest of $\tau$'s is due to the quantity $(\mathcal{T} - \tau)$ approaching zero (where $\mathcal{T}$ is the number of timesteps in the trajectory). These autocorrelation functions were subsequently Fourier transformed to produce the spectra presented in the text.}
\label{supp_fig:superML_cfxns}
\end{figure*}

\vfill

\pagebreak[4]

\section{Monolayer Dipole-Dipole Fluctuation Spectra}

\begin{figure}[h]
\centering
\includegraphics[width=2.5in]{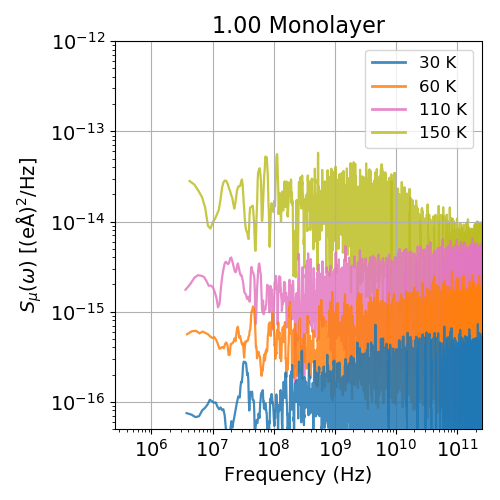}
\caption{$S_{\mu}(\omega)$ spectra of selected monolayer coverage runs. The spectra were smoothed over a window of five timesteps. Unlike the submonolayer spectra, monolayer spectra have no low frequency features at any temperature, although the magnitude of their white noise increases with temperature.}
\label{supp_fig:ML_spectra}
\end{figure}


\section{Low Temperature Supermonlayer Dipole-Dipole Fluctuation Spectra}

\begin{figure}[h]
    \centering
    \includegraphics[width=2.5in]{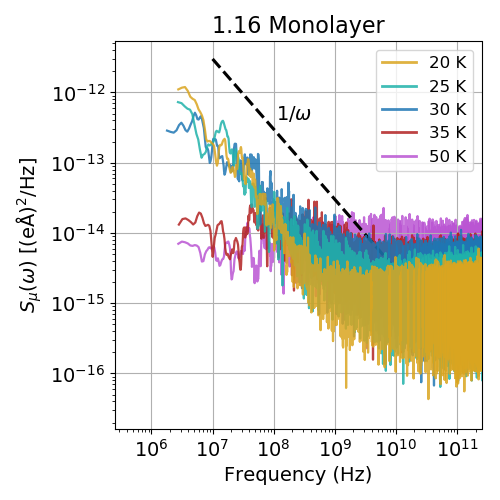}
    \caption{Selected low temperature $S_{\mu}(\omega)$ spectra at the 1.16 ML coverage. The spectra were smoothed over a window of five timesteps.}
    \label{supp_fig:superML_lowT}
\end{figure}

\vfill
\pagebreak[4]

\section{Plots of TLF System Behavior}

\begin{figure}[h]
    \centering
    \includegraphics[width=3in]{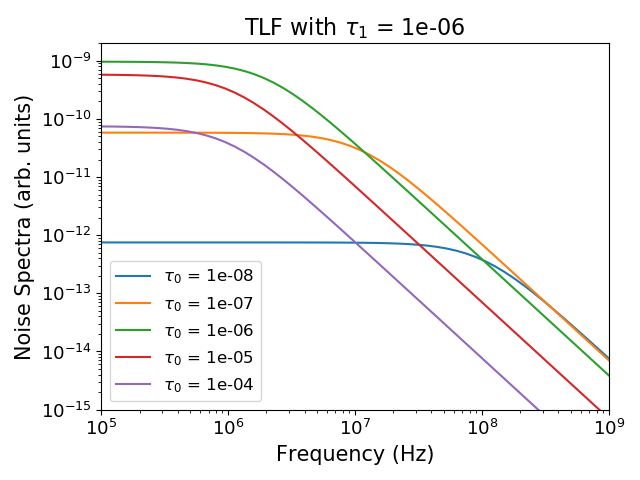}
    \caption{Examples of TLF noise spectra with $\tau_1$ fixed at 10 $\mu s$ and $\tau_0$ being varied (similar to Figure 4.6 in \textit{Zanola}). $\Delta I$ was set to 0.044, the same magnitude as the Layer 1$\leftrightarrow$2 transitions. TLF noise reaches a maximum when $\tau_0=\tau_1$. When $\tau_0>\tau_1$, the TLF noise decreases while the cutoff frequency ($\omega_{c} = (1/\tau_0 + 1/\tau_1)$) remains mostly unchanged. When $\tau_0<\tau_1$, the TLF noise decreases and $\omega_{c}$ increases, effectively shifting the frequency curve to the right.}
    \label{supp_fig:tlf_tau_variation}
\end{figure}

\begin{figure}[h]
    \centering
    \includegraphics[width=5.237in]{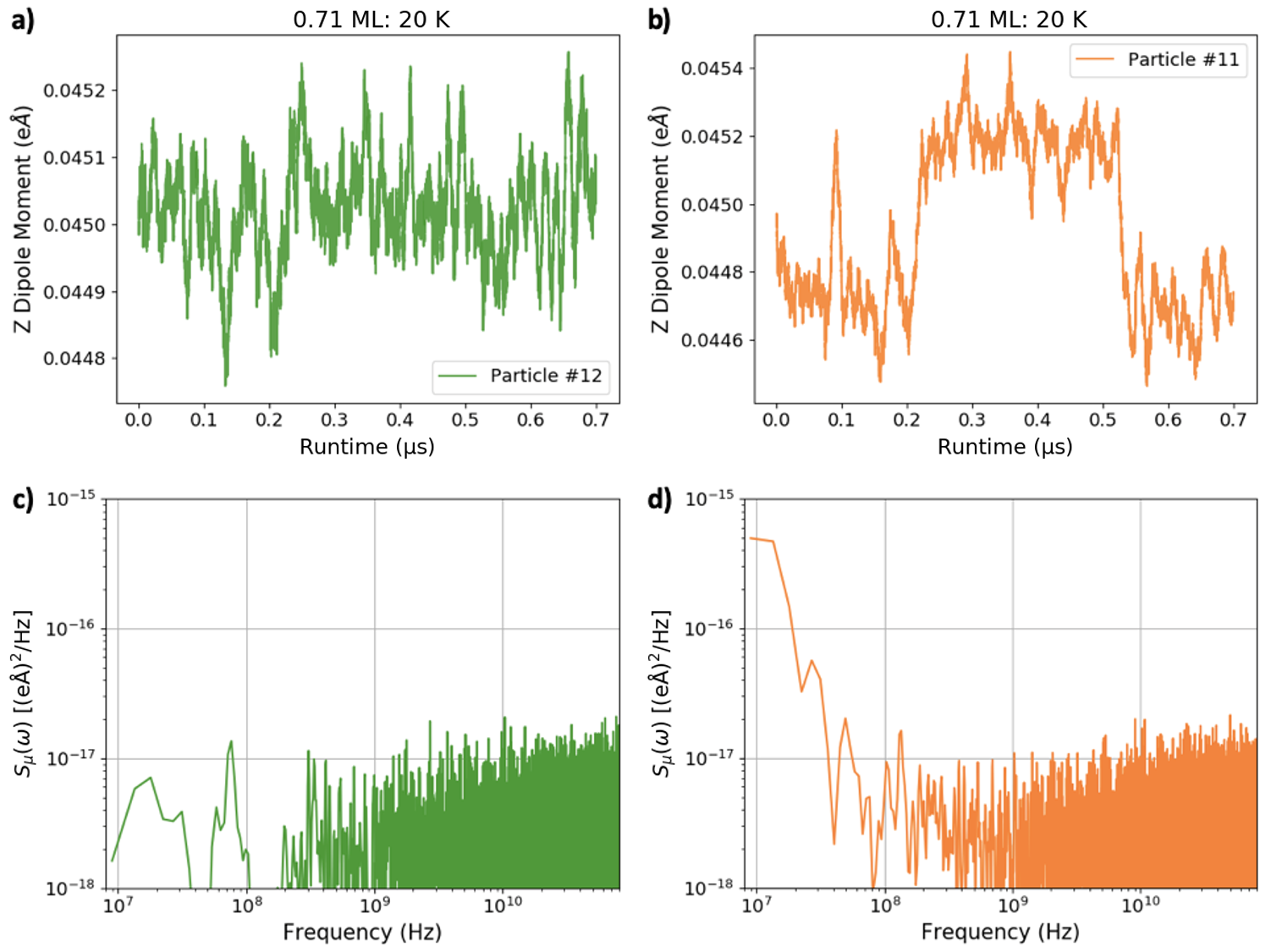}
    \caption{Emergence of low frequency $S_{\mu}(\omega)$ noise in a 0.7 $\mu$s stretch of the 0.71 ML submonolayer coverage trajectory run at 20 K. The top panels show the 10,000-frame moving average of the $z$ dipole moment for two different adsorbates: \textbf{(a)} Particle \#12, which shows deviations from a single value of $~4.5\times10^{-2}$ e\r{A}, and \textbf{(b)} Particle \#11, which shows two distinct dipole regions separated by $5\times10^{-4}$ e\r{A}. The bottom panels show the corresponding FT spectra for each particle: \textbf{(c)} Because Particle \#12 never fluctuates between discrete average dipole regions, its spectrum is featureless. \textbf{(d)} Because Particle \#11 fluctuates between two dipole regions, its spectrum has the TLF scaling of $1/\omega^2$.}
    \label{supp_fig:subML_lowfreq_noise}
\end{figure}

\begin{figure}[h]
    \centering
    \includegraphics[width=3in]{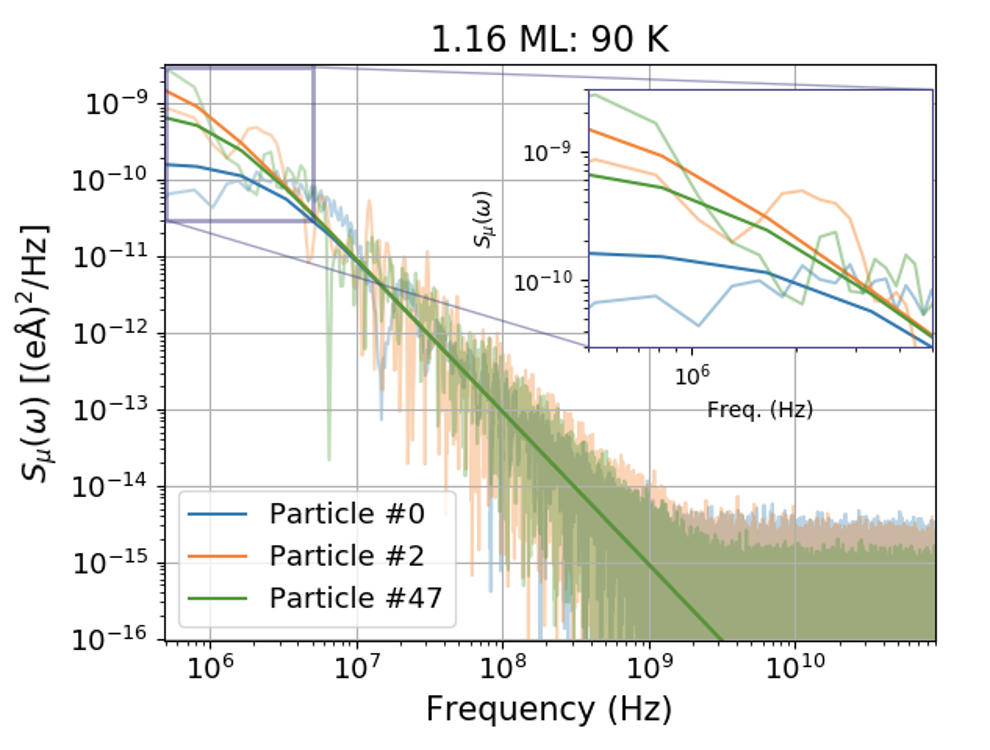}
    \caption{Comparison of simulated spectra (lighter shade) and analytical spectra calculated from the two-level fluctuator model (darker shade) for selected particles of the 1.16 ML run at 90 K.}
    \label{supp_fig:analytical_form_vs_sim}
\end{figure}
